\documentclass[12pt,preprint]{aastex}
\begin{document}
\title{Current Star Formation in the Perseus Molecular Cloud: \\ Constraints from Unbiased Submillimeter and Mid-Infrared Surveys}
\author{Jes K. J{\o}rgensen\altaffilmark{1}, Doug Johnstone\altaffilmark{2,3}, Helen Kirk\altaffilmark{3,2}, \& Philip C. Myers\altaffilmark{1}}
\altaffiltext{1}{Harvard-Smithsonian Center for Astrophysics, 60 Garden Street MS42, Cambridge, MA 02138, USA {\tt jjorgensen@cfa.harvard.edu}}
\altaffiltext{2}{Herzberg Institute of Astrophysics, National Research Council of Canada, 5071 West Saanich Road, Victoria, BC V9E 2E7, Canada}
\altaffiltext{3}{Department of Physics \& Astronomy, University of Victoria, Victoria, BC, V8P 1A1, Canada}

\begin{abstract}
  We present a census of the population of deeply embedded young
  stellar objects (YSOs) in the Perseus molecular cloud complex based
  on a combination of Spitzer Space Telescope mid-infrared data from
  the ``Cores to Disks'' (c2d) legacy team and JCMT/SCUBA
  submillimeter maps from the COMPLETE team. The mid-infrared sources
  detected at 24~$\mu$m and having $[3.6]-[4.5]>1$ are located close
  to the center of the SCUBA cores, typically within 15\arcsec\ of
  their peaks. The narrowness of the spatial distribution of
  mid-infrared sources around the peaks of the SCUBA cores suggests
  that no significant dispersal of the newly formed YSOs has
  occurred. This argues against the suggestion that motions of
  protostars regulate the time scales over which significant
  (Bondi-Hoyle) accretion can occur. The YSOs are found to have red
  $[3.6]-[4.5]$ and $[8.0]-[24]$ colors, but not comparable red
  $[5.8]-[8.0]$ colors. The most deeply embedded YSOs are found in
  regions with high extinction, $A_V \ge 5$, similar to the extinction
  threshold observed for the SCUBA cores. All the SCUBA cores with
  high concentrations have embedded YSOs, but not all cores with low
  concentrations are starless. From the above considerations a
  relatively unbiased sample of 49 deeply embedded YSOs is
  constructed. Embedded YSOs are found in 40 of the 72 SCUBA cores
  with only three cores harboring multiple embedded YSOs within
  15$''$. The equal number of SCUBA cores with and without embedded
  YSOs suggests that the time scale for the evolution through the
  dense prestellar stages, where the cores are recognized in the
  submillimeter maps and have central densities of $5\times
  10^4-1\times 10^5$~cm$^{-3}$, is similar to the time scale for the
  embedded protostellar stages. The current star formation efficiency
  of cores is estimated to be approximately 10--15\%. In contrast, the
  star formation efficiency averaged over the cloud life time and
  compared to the total cloud mass is only a few percent, reflecting
  also the efficiency in assembling cloud material into the dense
  cores actually forming stars.
\end{abstract}


\section{Introduction}
Any theory of low-mass star formation should not only be able to make
predictions for one characteristic stage such as the core mass
spectrum in molecular clouds or the initial mass function of emerging
stars - but also relate those through both pre- and protostellar
stages. A number of the very basic questions of star formation still
remain unanswered and debated (see, e.g., \cite{difrancescoppv},
\cite{wardthompsonppv} and \cite{ballesterosppv} for recent reviews
and discussions of both observational and theoretical studies), which
can be addressed with large homogeneous surveys from mid-infrared
through (sub)millimeter wavelengths and well-defined lists of sources
such as those presented in this paper. For example, is the star
formation process highly dynamical with cores being transient
phenomena in a turbulent medium and are protostars only accreting in
short periods of time when ``passing by'' these dense environments
(i.e., Bondi-Hoyle accretion)? Or is low-mass star formation a more
slowly evolving process as preferred in traditional ambipolar
diffusion scenarios and in that case, how important are other effects
such as turbulence in regulating the speed by which the process
occurs? How efficient is the star formation process in turning
prestellar dust and gas into young stars? Are there specific
relationships between the natal prestellar cores and the emerging
protostars, for example: do the properties of prestellar cores reflect
in whether a given core actually forms a star and perhaps whether it
forms a single or multiple system? Many of these discussions hinge on
statistical arguments, for example estimates of the evolutionary rate
of protostars through counts of the number of YSOs in different
stages. Previous studies of the deeply embedded stages were hampered
by low sensitivity single element (sub)millimeter receivers and
confusion in low resolution, low sensitivity infrared observations
such as those from the IRAS and ISO satellites. In the last few years
systematic, detailed surveys of larger regions have been made possible
using wide area imaging at high resolution and sensitivity using
submillimeter telescopes such as the JCMT with SCUBA and mid-infrared
telescopes including the Spitzer Space Telescope and its Infrared
Array Camera (IRAC) and Multiband Imaging Photometer (MIPS). In this
paper we present a census of the deeply embedded low-mass protostars
in the Perseus molecular cloud utilizing recently obtained large scale
Spitzer/c2d \citep{perspitz,rebull06} and JCMT/SCUBA maps
\citep{kirk06} and discuss some of the strong constraints for theories
of low-mass star formation emerging from these homogeneous datasets.

The earliest, most deeply embedded stages of low-mass protostars or
the latest stages of prestellar cores close to the onset of collapse
are important for insight into the physical conditions regulating
low-mass star formation. Approaching the problem from the prestellar
stages, continuum observations provide strong constraints on the
distribution of dust, and through inference the gas, in star forming
regions. In particular, large scale mapping with submillimeter and
millimeter wavelength continuum bolometers
\citep[e.g.,][]{motte98,johnstone00,johnstone04,
  hatchell05,enoch06,kirk06,young06} have enabled relatively unbiased
tabulations of dust condensations that either just have or are likely
to form protostars. The distribution of dust emission can also be used
to infer the column density distribution of the dust in both pre- and
protostellar cores and from there potentially derive self-consistent
density and temperature distributions through detailed radiative
transfer modeling
\citep[e.g.,][]{hogerheijde00sandell,evans01,jorgensen02,shirley02}. Alternatively,
observed parameters such as peak flux or size can be used as empirical
diagnostics to estimate which cores will undergo collapse
\citep[e.g.,][]{johnstone00,kirk06}.

The dust continuum maps themselves do not answer the question whether
a given core already has formed a central protostar: for that deep
mid-infrared observations are needed. Furthermore mid-infrared
observations are important as an evolutionary indicator once the
protostar itself has formed: its infrared excess, spectral slope in
the mid-infrared range or combination of colors are typically used to
distinguish between objects in different stages
\citep[e.g.,][]{lada87,greene94,allen04}. Still, the infrared colors
themselves do not provide an unambiguous mapping to object type and/or
evolutionary status. Deeply embedded YSOs for example share many of
the same SED characteristics as more evolved edge-on disk
systems. Also, since the most deeply embedded YSOs emit a significant
fraction of their luminosity at far-infrared and submillimeter
wavelengths (Class 0 objects are for example observationally defined
as objects emitting more than 0.5\% of their total luminosity at
wavelengths longer than 350~$\mu$m; e.g., \cite{andre93,andreppiv})
establishing their full SEDs through these wavelengths is
important. All of these issues illustrate the importance of combining
submillimeter and mid-infrared observations as is done in this paper.

The Spitzer Space Telescope ``From Molecular Cores to Planet Forming
Disks (Cores to Disks; c2d)'' legacy team \citep{evans03} has surveyed
5 of the nearby star forming clouds using the mid-infrared cameras,
IRAC and MIPS, with the aim of characterizing the ongoing star
formation in each. The ``Coordinated Molecular Probe Line Extinction
Thermal Emission (COMPLETE)'' survey of star forming regions
\citep{goodman04,ridge06} collect data for the three northern of these
regions in a range of molecular lines, extinction and dust
continuum. In this paper we combine the information from the c2d and
COMPLETE surveys of Perseus\footnote{Throughout this paper we follow
  the Spitzer/c2d papers and adopt a distance of 250~pc to Perseus
  (see discussion in \cite{enoch06}).}. Perseus is a good test case
for such a comparison with a large number of YSOs in different
environments from large clusters such as NGC~1333 and IC~348 to
objects in relative isolation or in small groups like L1448, L1455,
Barnard~1 and Barnard~5. This paper builds on the analysis of Perseus
through Spitzer \citep{perspitz,rebull06} and JCMT/SCUBA observations
\citep{kirk06} by examining the relation between SCUBA cores and
Spitzer sources and thereby building a sample of deeply embedded
objects. The paper is laid out as follows: \S\ref{cluster} first
examines the general association of mid-infrared sources and the SCUBA
cores, the mid-infrared colors of the sources found to be embedded in
SCUBA cores (\S\ref{midir}) and the properties of the SCUBA cores with
embedded mid-infrared sources (\S\ref{submm}). \S\ref{overallcloud}
examines the properties of the embedded sources compared to the
overall cloud environment in terms of extinction
(\S\ref{extincttracer}) and submillimeter flux
(\S\ref{submmtracer}). The results of \S\ref{cluster} and
\S\ref{overallcloud} are combined and an unbiased sample of deeply
embedded protostars in Perseus is established. \S\ref{discuss}
presents the sample of sources (\S\ref{sample}) and points to a number
of important observational constraints for models of star formation,
in particular in terms of dispersal of newly formed protostars
(\S\ref{dispersal}), time scales for the evolution through the
prestellar stages (\S\ref{timescales}) and the star formation
efficiency (\S\ref{SFE}). The results in this paper will also serve as
an important path-finder for comparisons to other clouds surveyed at
both mid-infrared and submillimeter wavelengths.

\section{Mid-infrared sources associated with dense cores}\label{cluster}
The first task in building up an unbiased list of embedded YSOs is to
define what is meant by an association between a mid-infrared source
and a submillimeter core. For this we first explore the distribution
of mid-infrared Spitzer sources in comparison to the dust
condensations seen in the SCUBA maps. In this section we consider the
distribution and properties of the nearest MIPS 24~$\mu$m source to
each submillimeter core and use these to draw general conclusions
about the properties of the embedded mid-infrared sources. In
\S\ref{submmtracer} we turn the table and consider the properties of
all the MIPS 24~$\mu$m sources in the catalog.

Deeply embedded YSOs are expected to have a steeply rising SED from
the IRAC through MIPS wavelengths due to the cold envelope emitting at
longer wavelengths while absorbing and reprocessing emission from the
central star+disk system at shorter wavelengths. We therefore expect
any deeply embedded YSO detected at the IRAC wavelengths also to be
detected at 24~$\mu$m whereas the inverse may not be true. The Class 0
protostar IRAS~16293-2422 is for example seen in IRS observations at
23--35~$\mu$m \citep{iras16293letter} (and likewise in more sensitive
MIPS observations) but not at the IRAC wavelengths. As a first try we
therefore examine the relationship between MIPS 24~$\mu$m sources from
the c2d observations \citep{rebull06} and the SCUBA cores from the
list of \cite{kirk06} and H.~Kirk et al. (in prep.)\footnote{The list
  of SCUBA cores used in this paper is based on a new processing of
  the SCUBA maps from \cite{kirk06} with smaller (3\arcsec) pixel size
  (H.~Kirk et al. in prep.).}. The one place where this may be
problematic is in regions where the 24~$\mu$m maps are confused or
bright emission saturates the images. We will return to these issues
later in this section. Many of the studied sources are also detected
at longer wavelengths by MIPS at 70 and 160~$\mu$m where the SED of a
typical embedded protostar peaks. On the other hand, these longer
wavelength observations are less sensitive, more likely to be
saturated and provide a lower angular resolution, and are therefore
less useful for identifying individual objects in an unbiased sense
(as is the goal of this paper), especially in confused regions. For
establishing the full SEDs of individual objects these data are
important of course. In the following we will refer to the sources
detected at 24~$\mu$m with MIPS as ``MIPS sources''.

One concern is how big the contamination of background sources is and
whether these are confused with the YSO population. Based on
comparison to the extra-galactic SWIRE data, \cite{rebull06} find that
most of the MIPS sources are background galaxies. Still, the surface
density of MIPS sources in the Perseus field is low enough that given a
random spatial distribution only 0.2 sources would be found within 
any 1~arcminute$^2$ - or likewise that
the probability of a chance alignment within 15\arcsec\ of a given
SCUBA core is just a few percent.

\begin{figure}
\resizebox{0.8\hsize}{!}{\includegraphics{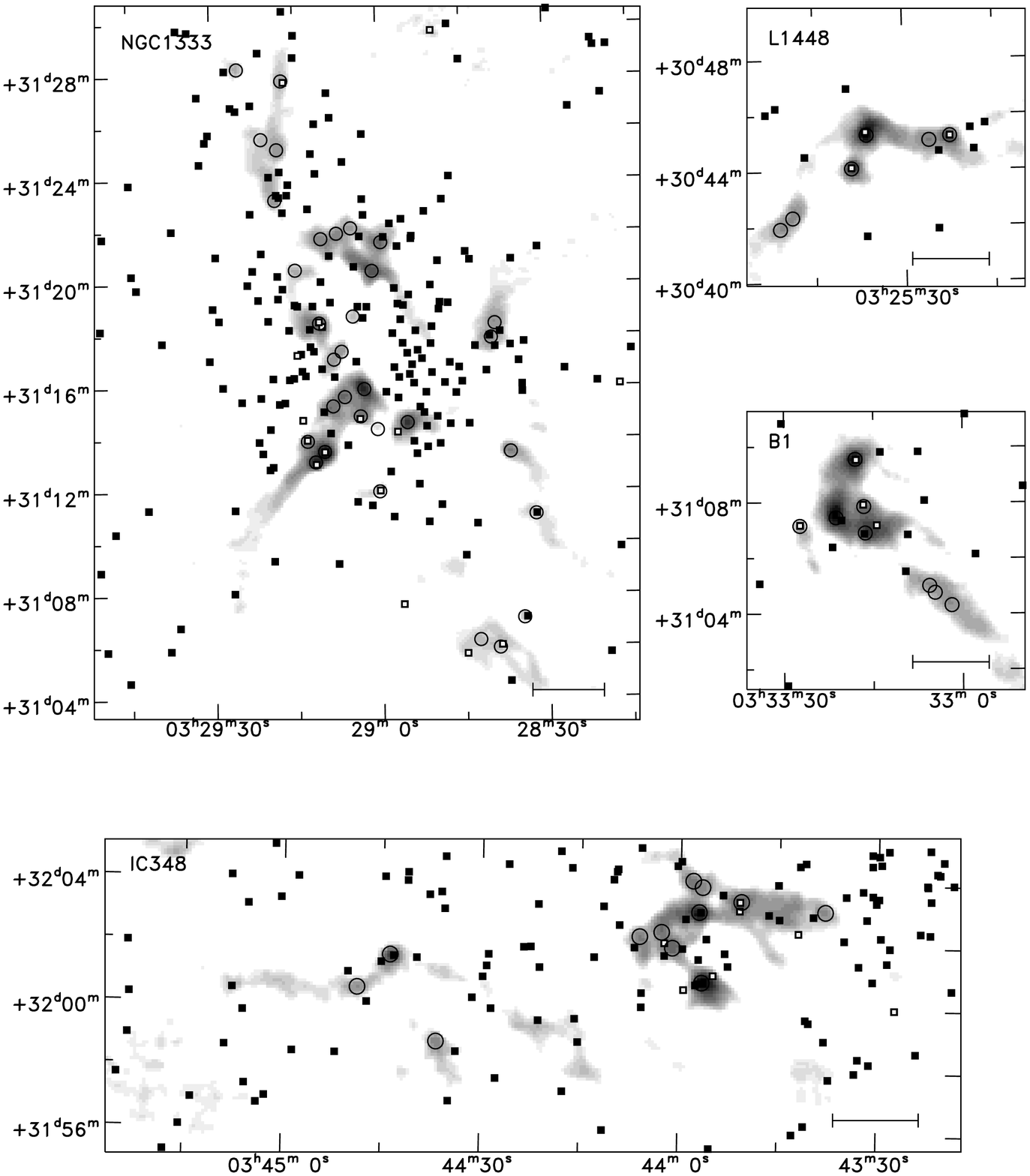}}
\caption{Distribution of SCUBA cores (circles) and MIPS sources
  (filled squares) plotted over SCUBA maps of four representative
  regions of Perseus from \cite{kirk06}. The sizes of the symbols of
  the SCUBA cores correspond to radii of 15$''$, which is the radius
  used for determining associations between SCUBA cores and
  mid-infrared sources. The MIPS sources with $[3.6]-[4.5] > 1.0$ and
  $[8.0]-[24] > 4.5$ are singled out with open squares. A distance of
  0.2~pc is indicated by the scale bar in the lower right corner of
  each panel.}\label{source_overlay}
\end{figure}

\begin{figure}
\resizebox{0.5\hsize}{!}{\includegraphics{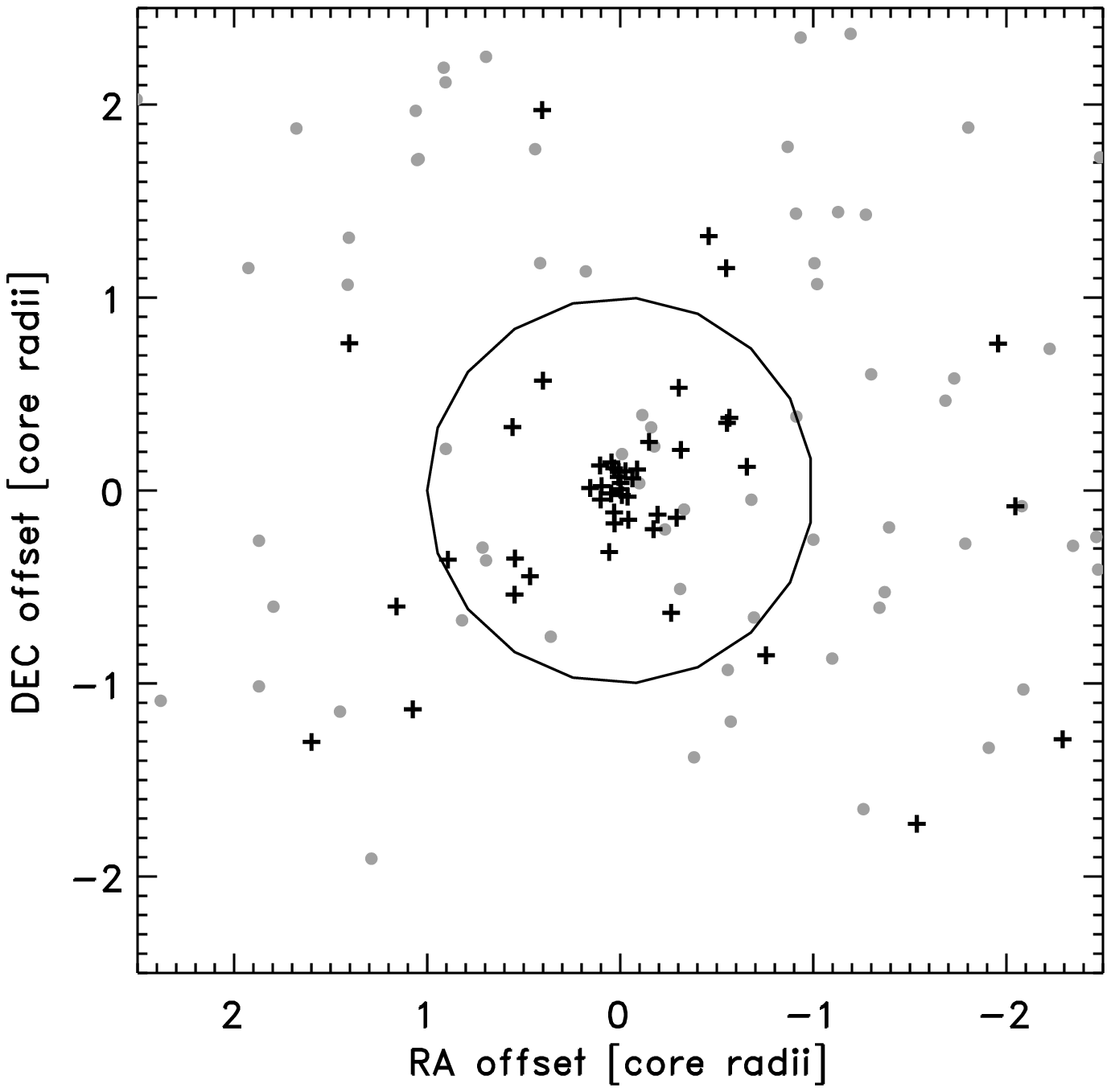}}

\resizebox{0.5\hsize}{!}{\includegraphics{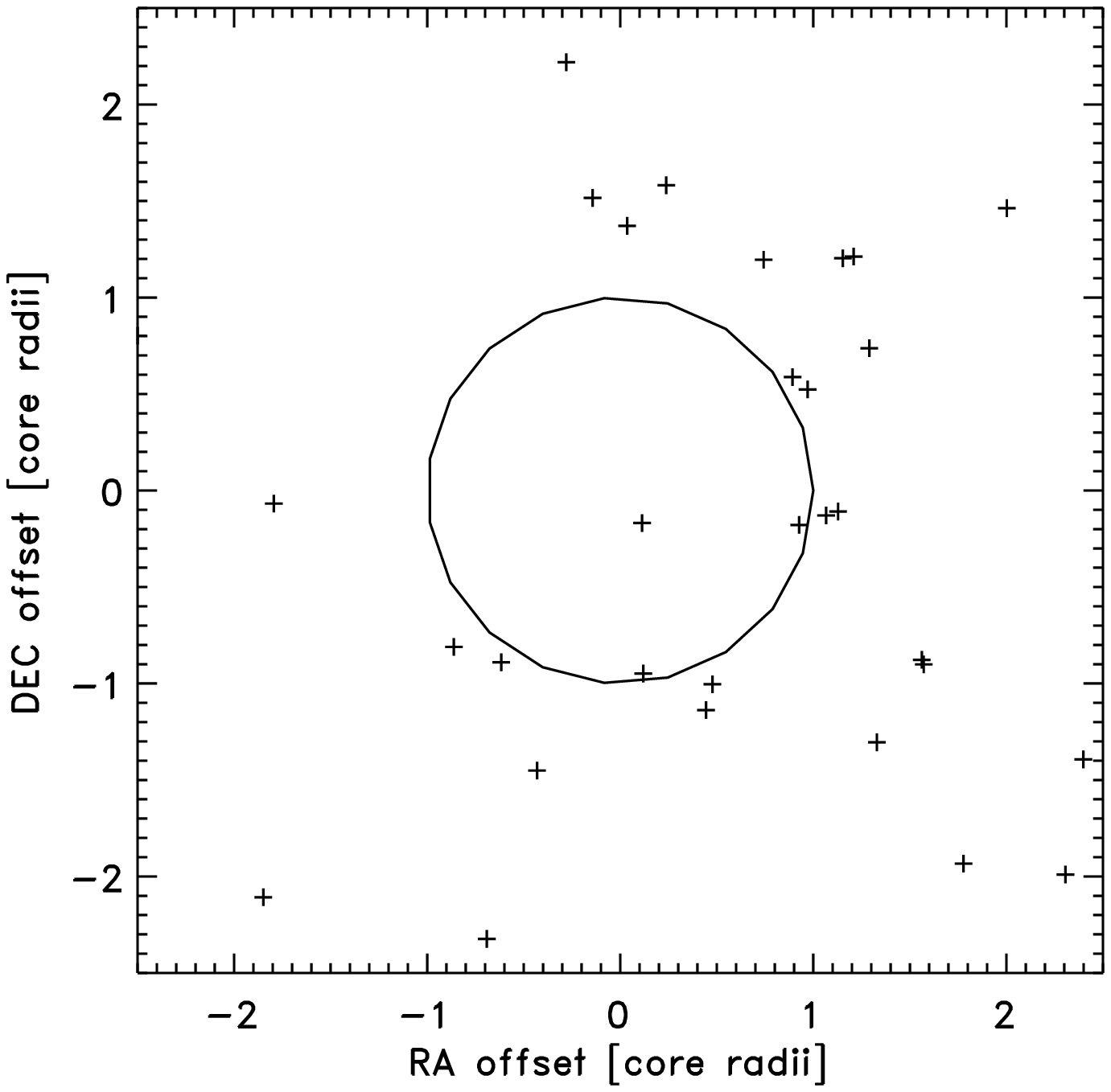}}
\caption{Distribution of MIPS sources with IRAC detections around the
  center of the nearest SCUBA core (all shifted to a common center)
  for any given MIPS source (\emph{upper panel}). The distances are
  given in units of the core radii. Sources with $[3.6]-[4.5] \ge 1.0$
  are shown with black plus signs and sources with $[3.6]-[4.5] < 1.0$
  grey dots. For comparison the bottom panel shows the similar plot
  for a random distribution of the MIPS sources over the SCUBA
  field.\label{clustdia}}
\end{figure}

\begin{figure}
\resizebox{\hsize}{!}{\includegraphics{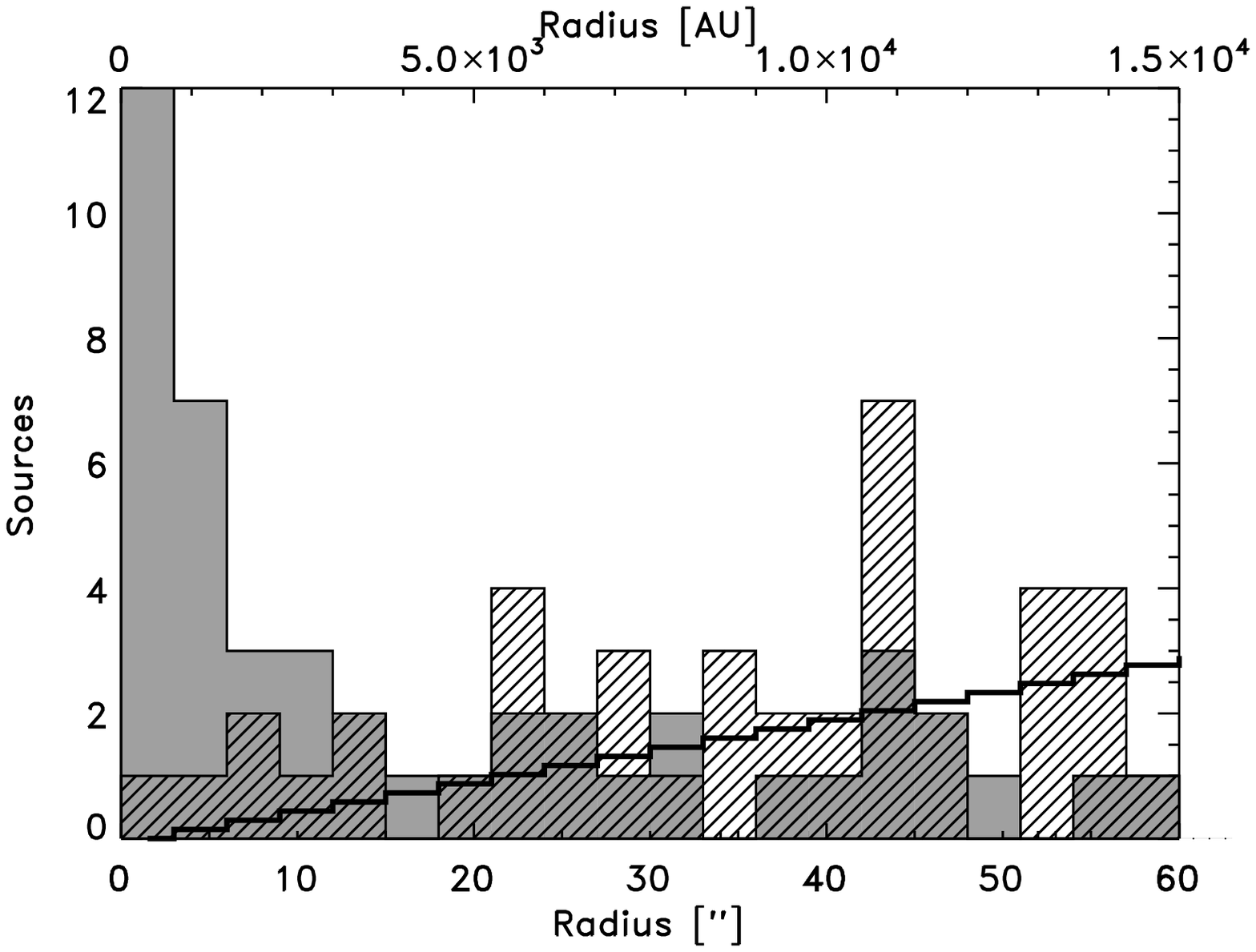}}
\caption{Distribution of number of mid-infrared sources as a function of
  distance. The filled grey histogram indicates the distribution of
  sources with $[3.6]-[4.5] \ge 1.0$ and the hatched histogram the
  distribution of sources with $[3.6]-[4.5] < 1.0$. The thick black
  line indicates the prediction from a uniform source distribution
  with the same surface density as the observed
  distribution.\label{clustdia_hist}}
\end{figure}

Fig.~\ref{source_overlay} shows the distribution of mid-infrared MIPS
sources and submillimeter SCUBA cores overlaid on SCUBA maps in a
few of prominent regions of Perseus. A number of the SCUBA cores
are associated with mid-infrared sources within rather small
distances. Significant clustering is also seen in other manners, for
example the clustering of the submillimeter cores or the mid-infrared
sources by themselves. In this paper we only focus on the association
of the mid-infrared sources with the submillimeter cores. Of the 72
cores from the SCUBA map, 36 (50\%) are found to have a MIPS
counterpart within 15\arcsec\ of the peak (31, or 43\%, have a MIPS
source within 10\arcsec). At the resolution of the Spitzer/MIPS
24~$\mu$m and SCUBA observations (approximately 6$''$ and 15$''$,
respectively) it appears that most of the SCUBA cores are associated
with one MIPS source only (Table~\ref{multiplicity}).
\begin{table*}
\caption{Number of cores with a given number of (specified) sources within a specific radius.}\label{multiplicity}
\begin{center}
\begin{tabular}{lllll}\hline\hline
                                        & \multicolumn{4}{c}{Number of SCUBA cores with}\\
                                        & 0  sources & 1  source & 2  sources & 3 sources \\ \hline
MIPS source within 15$''$               & 36\tablenotemark{a} & 33 & 3  & 0 \\
MIPS source within 30$''$               & 28\tablenotemark{a} & 34 & 7  & 3 \\
``red''\tablenotemark{b} MIPS source within 15$''$ & 45 & 26 & 1  & 0 \\
``red''\tablenotemark{b} MIPS source within 30$''$ & 42 & 26 & 3  & 1 \\ \hline
\end{tabular}
\end{center}
\tablenotetext{a}{As discussed in \S\ref{submm} four of these cores are associated with saturated MIPS sources not included in the c2d catalog.}
\tablenotetext{a}{Source with $[3.6]-[4.5] > 1.0$ and $[8.0]-[24] > 4.5$.}
\end{table*}
\cite{kirk06} defined the radius of each SCUBA core as $R_{\rm
  obs}=\sqrt{A/\pi}$ where $A$ is the area of the core at the lowest
contour identified by the \emph{Clumpfind} algorithm
\citep{williams94}. The typical core radii defined in this way range
from approximately 15--45$''$ (3,750--11,250~AU). Fig.~\ref{clustdia}
shows the distribution of MIPS sources around the center of any given
core in units of the radius of each core defined by
\citeauthor{kirk06} - and Fig.~\ref{clustdia_hist} illustrates the
same relationship as a one dimensional distribution as a function of
radius. The red mid-infrared sources are located close to the centers
of the submillimeter cores, typically within about 0.5 core radii
(significantly more peaked than a random distribution of sources) with
a distribution with a full width half maximum of a few tenths of a
core radius (or $\approx$~10\arcsec).  In \S\ref{submmtracer} we
discuss the properties of the red mid-infrared sources with larger
separations.
There is a caveat worth mentioning here: is the nearest submillimeter
core to each YSO really its parental core? Three effects could come
into play: (1) projection effects might play a role with some SCUBA cores
and mid-infrared sources simply being chance alignments, (2) the
motion of the YSOs with respect to their surroundings could be
significant making it possible for sources to travel to the vicinity
of different clumps, and (3) the parental core of a given YSO could
have been dissipated during the evolution of the YSO. The low surface
density of all the MIPS sources argue against the first being a
significant effect (see the lower panel of Fig.~\ref{clustdia}) while
the absence of sources with red colors, $[3.6]-[4.5] > 1$, outside the
cores and the narrowness of the distribution of their source-core
separations argue against the latter two points. In any case in the
above plots (and the discussion in the following sections) each
mid-infrared source is only counted once - even though a given
mid-infrared source in a few cases is the nearest neighbor to
multiple SCUBA cores.

\subsection{Properties of the MIPS sources}\label{midir}
\begin{figure}
\resizebox{0.8\hsize}{!}{\includegraphics{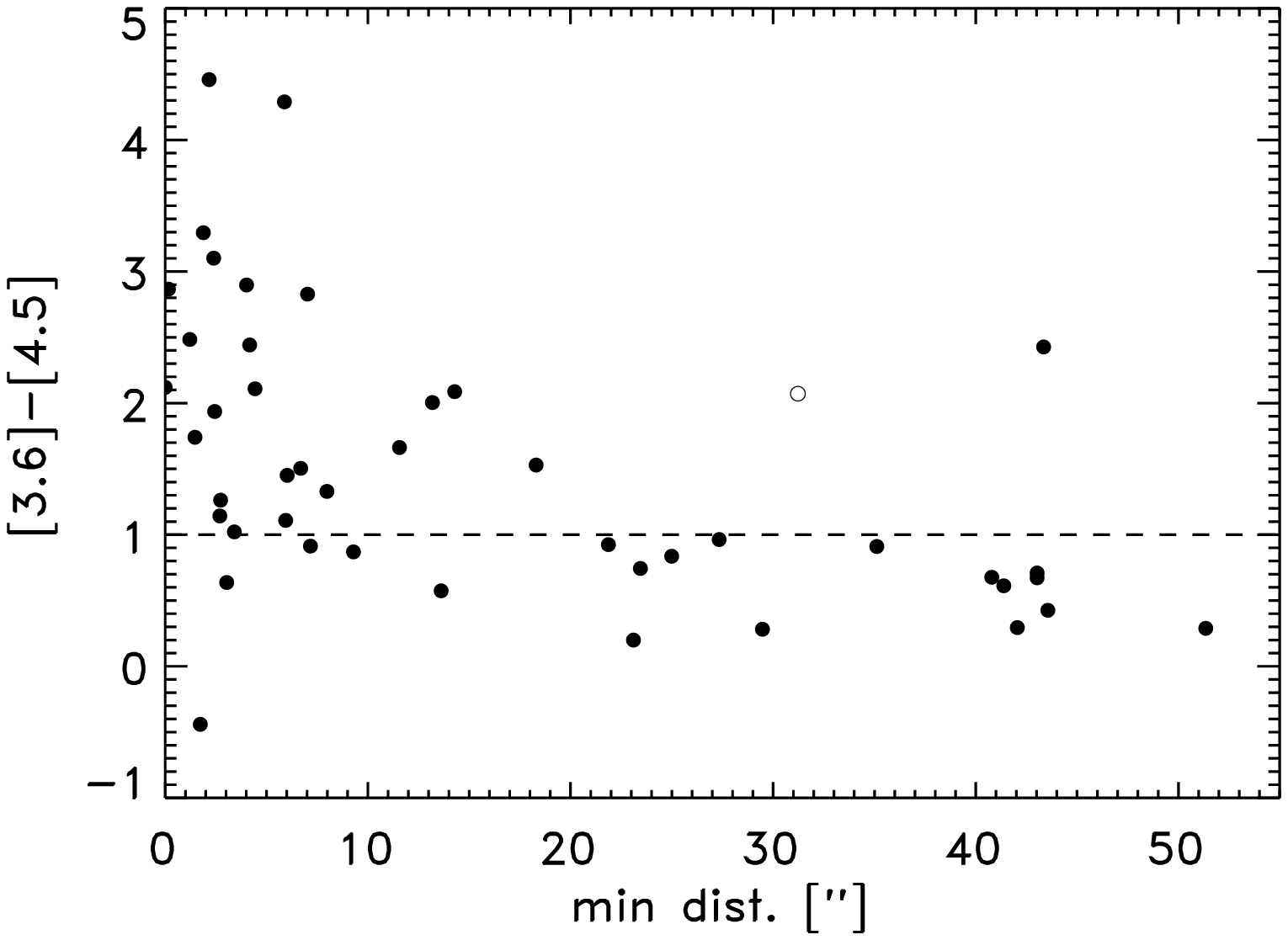}}
\resizebox{0.8\hsize}{!}{\includegraphics{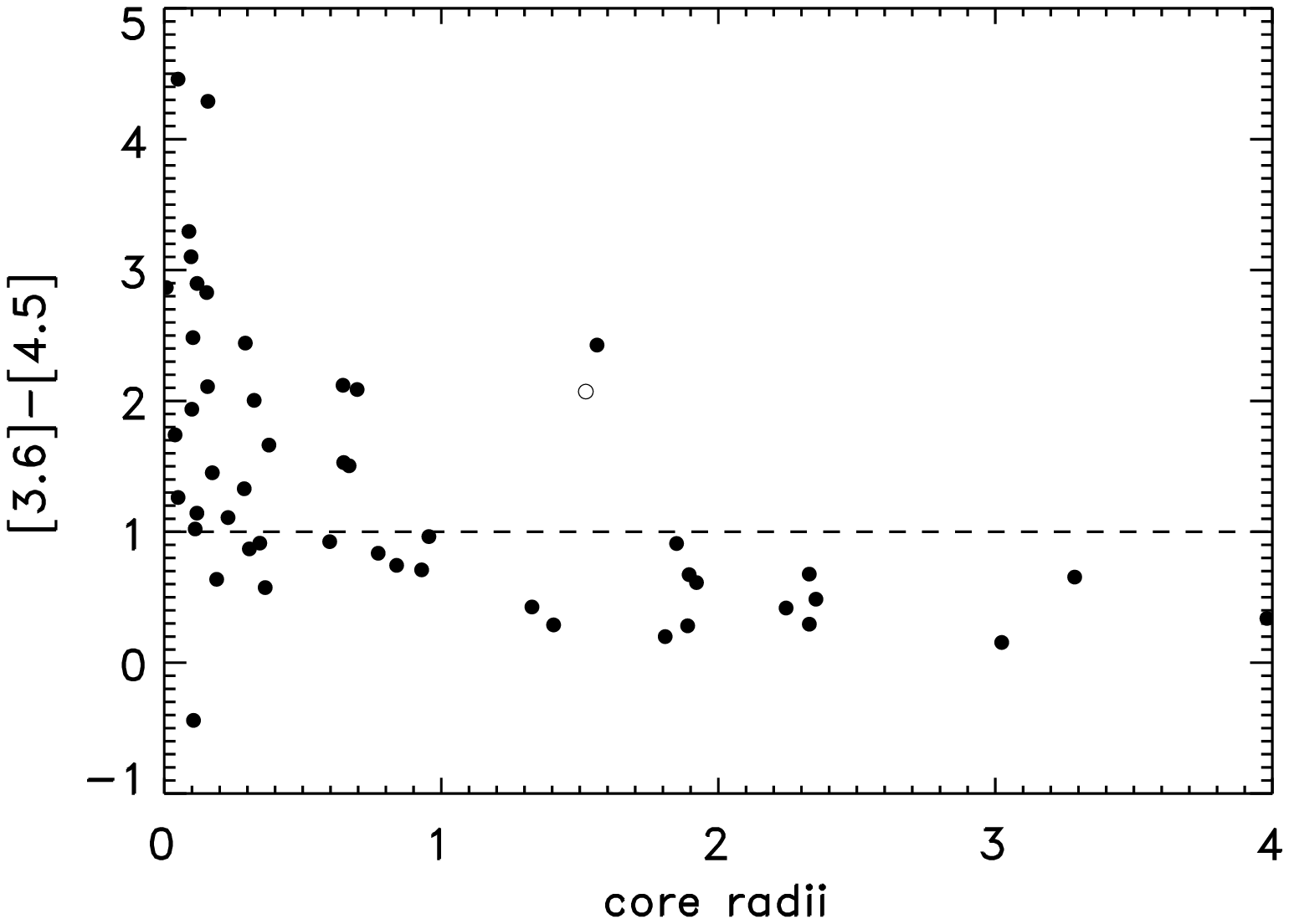}}
\caption{Nearest MIPS detection to each core and its $[3.6]-[4.5]$
  color vs. distance in arcseconds (upper panel) and core radii (lower
  panel). The shock associated with the L1448-IRS2 outflow is
  indicated by an open symbol.}\label{dist_i12}
\end{figure}

\begin{figure}
\plotone{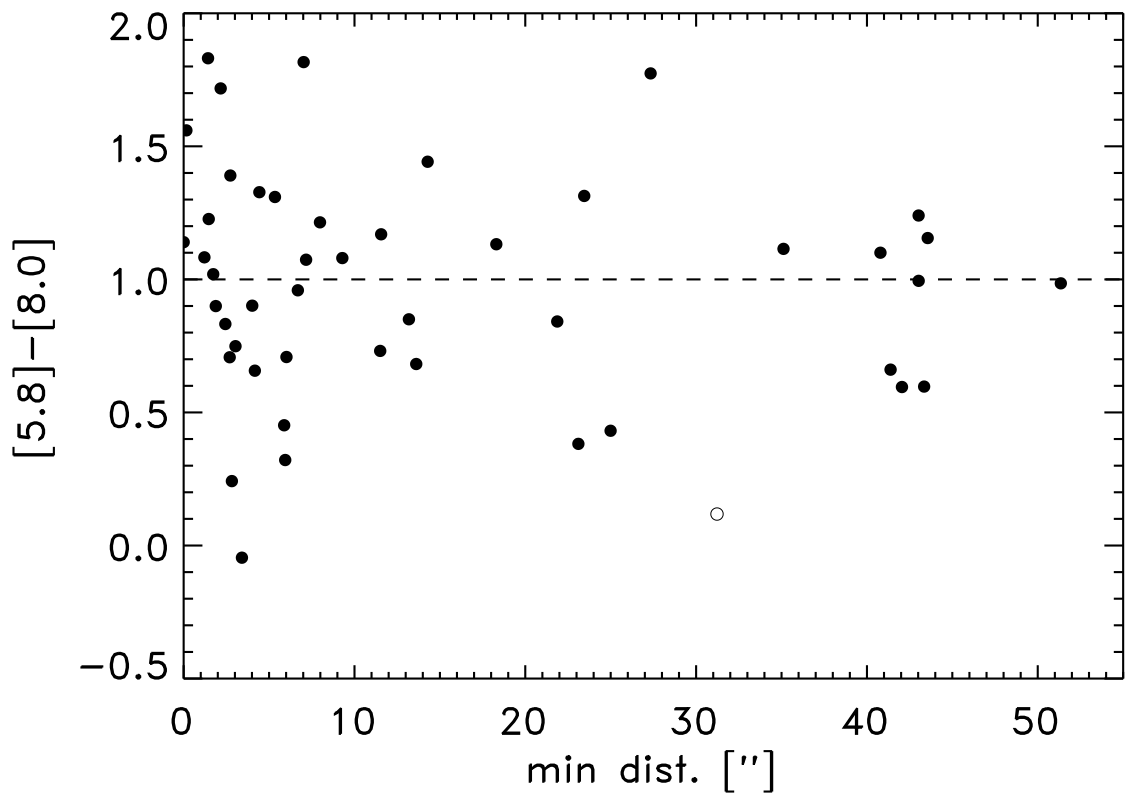}
\caption{Nearest MIPS detection to each core and its $[5.8]-[8.0]$
  color vs. distance.}\label{dist_i34}
\end{figure}

\begin{figure}
\plotone{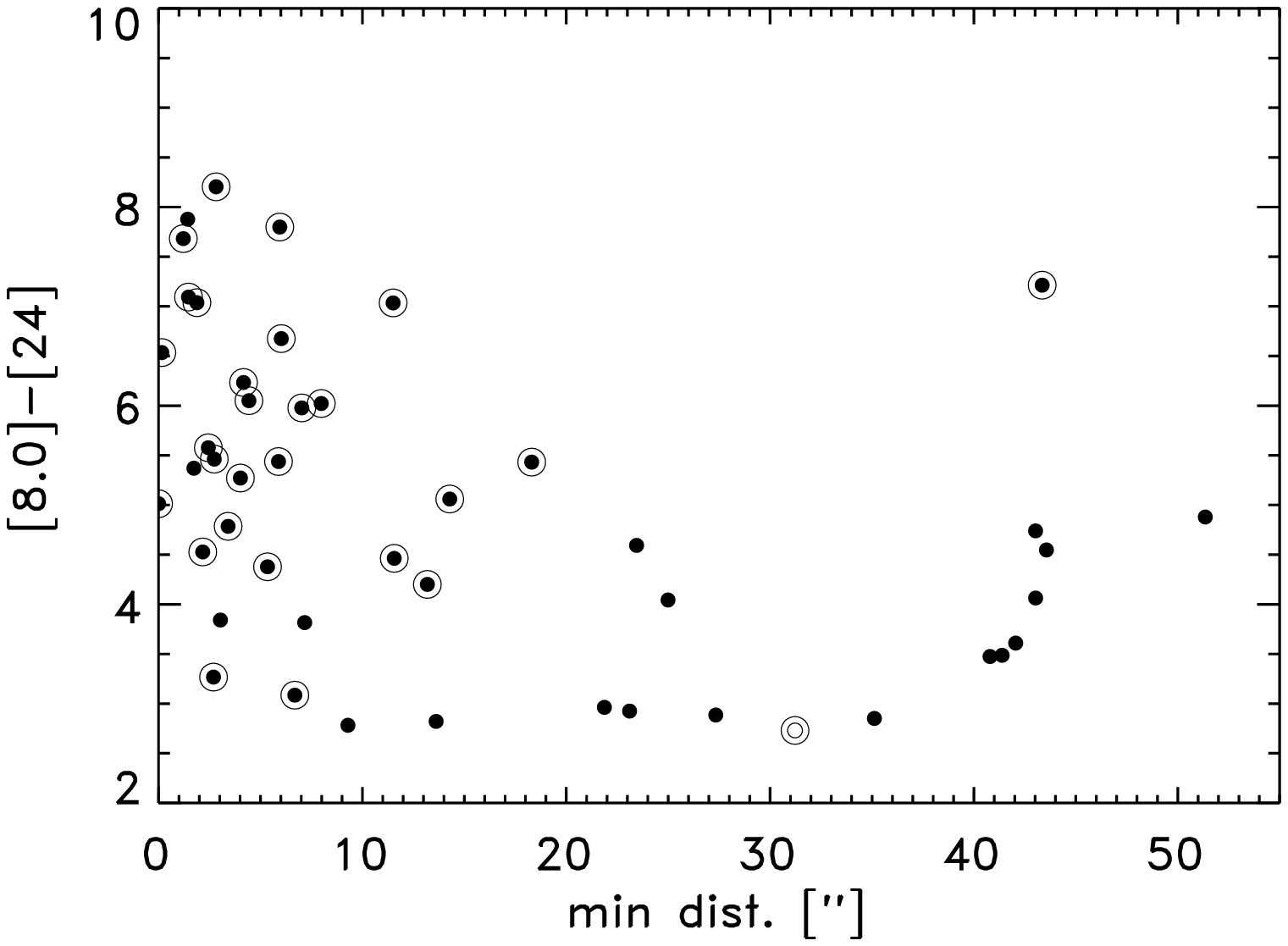}
\caption{Nearest MIPS detection to each core and its $[8.0]-[24]$
  color vs. distance. Sources with $[3.6]-[4.5] > 1$ are indicated
  with an additional outer circle around their symbol.}\label{dist_i45}
\end{figure}

Not only are the red MIPS sources located close to the center of the
submillimeter cores, but an interesting trend is also seen with their
colors. It is found that the sources within one core radius have red
$[3.6]-[4.5] \gtrsim 1$ colors. This is further illustrated in
Fig.~\ref{dist_i12} where the colors of the nearest MIPS source to
each core have been plotted as a function of distance. In contrast the
$[5.8]-[8.0]$ colors do not show a similar clear trend
(Fig.~\ref{dist_i34}). This is consistent with the results of
\cite{perspitz} who showed that the known Class 0 objects in Perseus
had very red $[3.6]-[4.5]$ colors but not similarly red $[5.8]-[8.0]$
colors. The very red $[3.6]-[4.5]$ colors imply that the central
protostars are significantly extincted by their ambient envelopes. It
is important to note that these objects in fact are detected at such
short mid-infrared wavelengths: this is in contrast to the traditional
picture of Class 0 objects not being detected at wavelengths shorter
than 10~$\mu$m. The $[5.8]-[8.0]$ colors in contrast are affected by a
number of effects, including absorption in the 10~$\mu$m silicate
feature, ice absorption and the intrinsic SED of the central star+disk
system \citep[see also discussion in][]{perspitz}. The $[8.0]-[24]$
color also shows an increase at small source-core separations
(Fig.~\ref{dist_i45}) and only few MIPS sources with $[8.0]-[24]
\gtrsim 4.5$ have distances larger than 15$''$. In contrast to the
comparison with the $[3.6]-[4.5]$ colors, the trend is not
unambiguous: some objects are found with small distances and
$[8.0]-[24] \lesssim 4.5$. This is similar to what was observed by
\cite{muzerolle04} who found that objects of differing embeddedness
(Class 0 through II) showed some overlap in their $[8.0]-[24]$
colors. One source (IRAS~03256+3055) has both very red $[3.6]-[4.5]$
and $[8.0]-[24]$ colors and is standing out a distance of about
43\arcsec\ to the nearest submillimeter core: inspection of the
submillimeter maps reveal clear extended dust continuum emission at
this position which is not identified as a core due to
confusion. Another red MIPS source with a large separation of 31$''$
is seen from visual inspection of the IRAC and MIPS images to be
located close to the L1448-IRS2 outflow. The MIPS source is likely to
be a chance alignment and its red $[3.6]-[4.5]$ colors just a result
of (unrelated) shocked H$_2$ emission in the L1448-IRS2 outflow,
showing up prominently at 4.5~$\mu$m. This source is excluded from the
sample in the following discussions.  

Each source in the c2d catalogs is classified with the aim of
separating stars, YSO candidates and other sources (mainly background
galaxies) according to cuts in the IRAC and MIPS color planes
\citep{delivery3}. In this way \cite{perspitz} identified 400
candidate YSOs in the 3.86~degree$^2$ of Perseus covered by the IRAC
observations. Not all the 39 MIPS sources (within 15$''$ of one of the
36 submillimeter cores with embedded YSOs) were included in that list,
however. As a first cut the high quality catalog used in
\cite{perspitz} exclude all sources that appear extended in one or
more IRAC bands as such sources have less reliable photometry. As
discussed in \citeauthor{perspitz} this is an issue for the embedded
YSOs, in particular, since they often appear extended in one or more
of the IRAC images, either due to H$_2$ emission from shocks in the
outflows or scattered light at the IRAC wavelengths. Of the 39 MIPS
sources discussed above, 18 show extended emission and are excluded
from the list of objects studied by \citeauthor{perspitz} 

In addition, the classification of YSOs described in \cite{delivery3}
relies on detection in at least three bands (of the four IRAC bands
and MIPS 24~$\mu$m) and very deeply embedded YSOs may therefore not be
picked out by these criteria if they for example are too faint at the
shortest IRAC wavelengths. Of the 39 MIPS sources discussed above, 29
(74\%) were classified as YSOs based on their colors. Of the remaining
objects, two are only detected at 24~$\mu$m but not the shorter IRAC
wavelengths, while the remaining eight are found to have red SEDs
(three times stronger MIPS flux than flux at nearest IRAC wavelength),
but do not fulfill the YSO criteria used in \cite{perspitz}. Of the 29
MIPS sources that are candidate YSOs based on their colors, 17 show
extended emission in one or more IRAC bands and only the remaining 12
are therefore included in the sample of YSOs in \cite{perspitz}. In
summary, the embedded YSOs from this paper should be added to the
general list of YSOs in Perseus from \citeauthor{perspitz} - about two
out of three of the embedded YSOs did not make it into that list.

\subsection{Properties of the cores}\label{submm}
\begin{figure}
\plotone{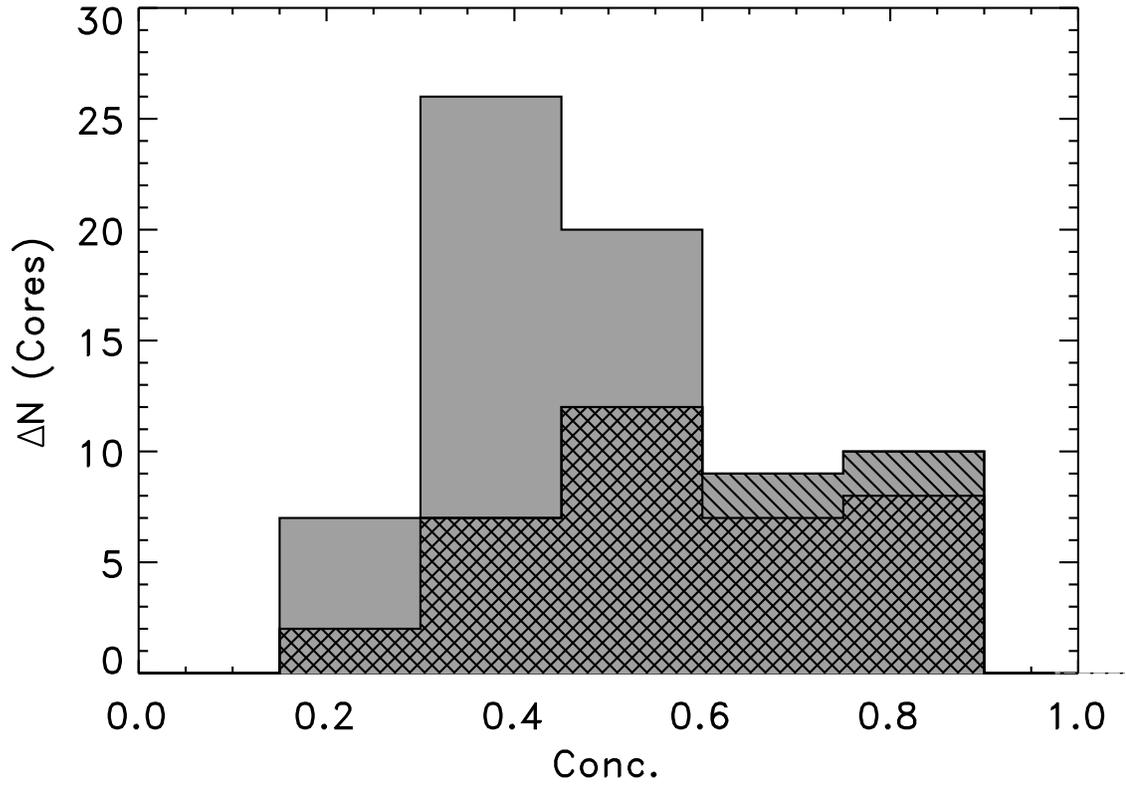}
\caption{Distribution of ``concentrations'' of SCUBA cores. Those with
  embedded YSOs within 15$''$ have been doubly-hatched while the
  additional 4 cores with concentrations higher than 0.6 discussed in
  the text have been singly-hatched.}\label{conc_histo}
\end{figure}

\begin{figure}
\plotone{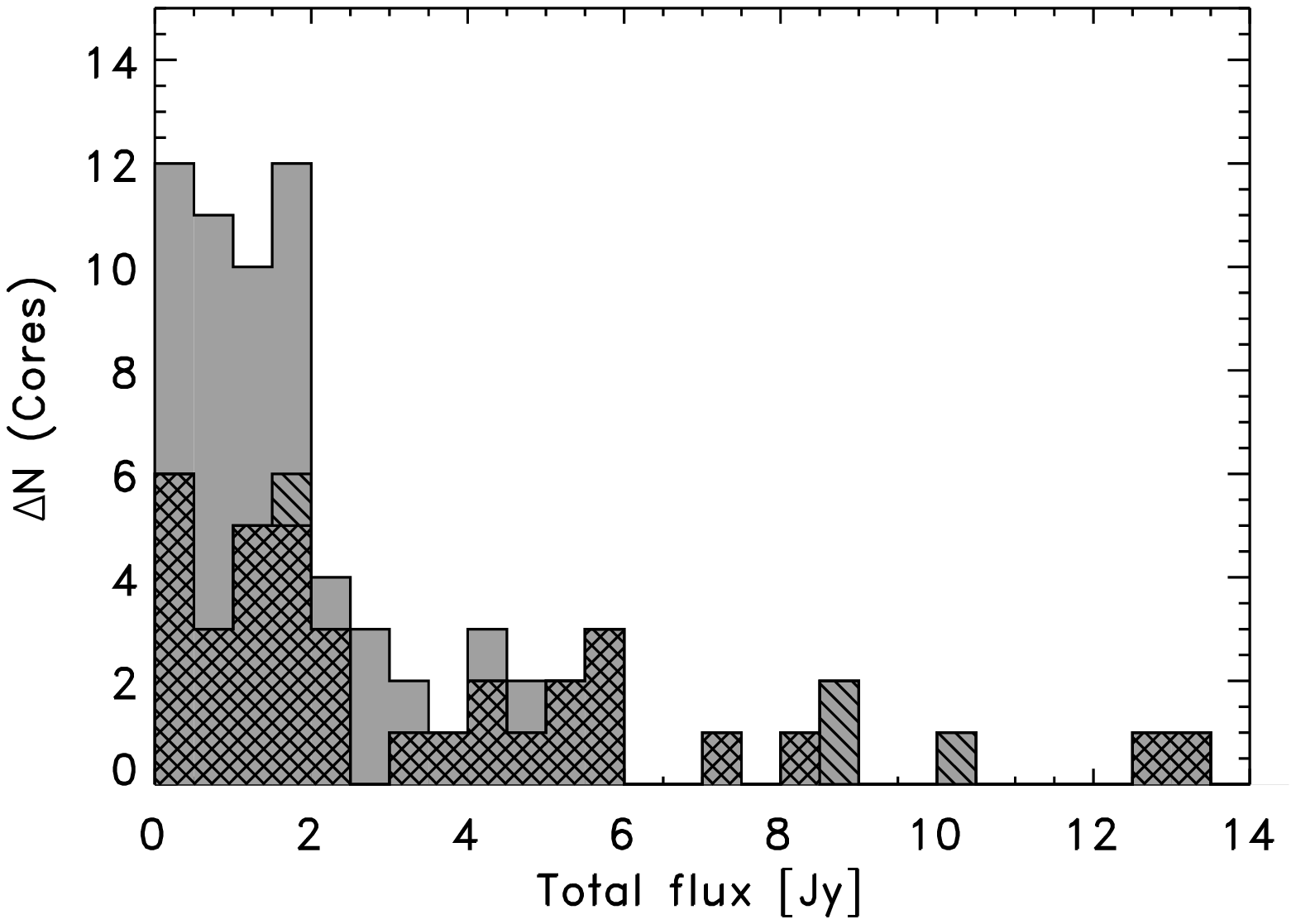}
\caption{As in Fig.~\ref{conc_histo} but for the distribution of total
  850~$\mu$m fluxes of the SCUBA cores.}\label{flux_histo}
\end{figure}

It is interesting to investigate whether the submillimeter cores with
embedded red sources have distinct properties compared to those
without, in particular to distinguish the starless from the star
forming cores based on the submillimeter maps by themselves. A
promising empirical characteristic of cores appear to be their
concentration:
\begin{equation}
C=1-\frac{1.13\, B^2S_{850}}{(\pi R^2_{\rm obs}f_0)}
\end{equation}
where $B$ is the beam size, $S_{850}$ the total flux from the SCUBA
observations at 850~$\mu$m, $R_{\rm obs}$ the measured radius and
$f_0$ the peak flux. From comparison between SCUBA maps and optical
and near-infrared images of shocks in the Barnard~1 region,
\cite{walawender05b1} found that all the SCUBA cores with
concentrations $>$~0.75 contained protostars with a decreasing
fraction with lower concentrations and none of the cores with
concentrations $<0.4$ having embedded protostars. \cite{walawender06}
and \cite{johnstone06} have extended this reasoning to IC~348 and
Orion A South, respectively. Fig.~\ref{conc_histo} and
\ref{flux_histo} compare the number of cores with a given
concentration and total 850~$\mu$m flux with embedded YSOs to the
total number of cores. There is a trend in that most cores with
concentrations higher than 0.6 have embedded YSOs. A number of effects
could cause this empirical distinction: as the concentration is a
measure of the brightness distribution of the submillimeter emission
it will be affected by the formation of a central protostar, i.e., an
object with an internal source of heating will show more emission
toward the center and therefore appear more concentrated. Also, just
considering prestellar cores: with the above definition Bonnor-Ebert
spheres stable against collapse have $C < 0.72$ \citep{johnstone00},
whereas cores with higher concentrations are either collapsing or
require support from other mechanisms (e.g., magnetic fields) in
addition to the thermal support. Still, even a highly centrally
condensed prestellar core such as L1544
\citep[e.g.,][]{evans01,crapsi05} has a concentration of $\approx$~0.5
estimated from SCUBA maps of the core and would therefore not fall
into the above group of objects.

There also seems to be a distinction between the SCUBA cores with and
without embedded YSOs in terms of their total submillimeter flux. We
find that 80\% of the starless cores have fluxes lower than 2~Jy (see
also Table~\ref{core_flux_table}) although the trend is not as clear
cut as for the concentrations. As for the concentrations, it is
interesting to note that a number of the cores with embedded YSOs do
have low fluxes. If the cores all had the same dust temperatures the
trend would imply an evolution with core mass. In cores with embedded
YSOs the dust temperature will be higher due to the heating by the
central protostar and the mass lower than the mass of a comparable
starless core with similar flux. Since most starless cores have fluxes
lower than 2--3~Jy (or masses lower than 1--1.5~$M_\odot$), this is
not a significant effect. No clear trend in any of these parameters
are seen for the cores with multiple embedded mid-infrared sources
(Table~\ref{multiplicity}) compared to the overall sample of cores
with only one embedded source.  
\begin{table}
\caption{Number of cores with and without embedded YSOs as a function of core flux.}\label{core_flux_table}
\begin{tabular}{llll}\hline\hline
Total 850~$\mu$m flux range [Jy]     & Total & With embedded YSOs & Mass range [$M_\odot$]\tablenotemark{a} \\ \hline
$<$~0.5   & 12 & 6 (50\%)      & $<$~0.25      \\
0.5--1    & 12 & 3 (25\%)      & 0.25--0.5 \\
1--2       & 21 & 11 (52\%)    & 0.5--1.0     \\
2--4       & 10 & 5 (50\%)      & 1.0--2.0     \\
4--6       & 10 & 8 (80\%)      & 2.0--3.0     \\
$>$~6      &  7  & 7 (100\%)    & $>$~3.0        \\ \hline
\end{tabular}

\tablenotetext{a}{Assuming a dust temperature of 15~K and a dust opacity of
0.0182~cm$^{2}$~g$^{-1}$ at 850~$\mu$m.}
\end{table}

Four SCUBA cores with concentrations larger than 0.6 do not have any
embedded MIPS sources from the criteria above. One example is
NGC~1333-IRAS2A, a well-studied Class 0 protostar which is clearly
identified but saturated in the MIPS 24~$\mu$m images. At the IRAC
wavelengths extended emission from its jet is seen, causing its
photometry to be unreliable similar to the problems of other Class 0
objects \citep{perspitz}. Furthermore its companion NGC~1333-IRAS2B
(separation of 30$''$) is clearly identified as a YSO at all four IRAC
bands and MIPS, but its SCUBA flux is dominated by the emission from
IRAS2A. Similar issues are also seen for the remaining SCUBA cores
with high concentrations, which are associated with the previously
known infrared sources, IRAS~03255+3103, NGC~1333-SVS12 and
-SVS13. None of the remaining starless cores (i.e., cores with no
associated mid-infrared sources) with concentrations less than 0.6 are
found to have embedded YSOs from visual inspection of the MIPS and
IRAC images.

Again, this poses the problem of whether there are more sources with
IRAC detections and red colors but no MIPS detections. Besides the
saturation issue of MIPS, this could also be the case in regions where
the sources are confused at the slightly worse MIPS resolution. Only
seven red sources are found with $[3.6]-[4.5]$ colors greater than 1.0
and no MIPS counterpart. These include the sources in the cores
discussed above, a shock knot associated with the NGC~1333-IRAS4A
outflow, three fainter companions to strong MIPS sources and a faint
(mJy level) IRAC source in a dust condensation southeast of L1448. So
generally, it appears that the MIPS detection requirement is useful to
identify embedded YSOs - although confusion might arise in regions of
many bright MIPS sources.

In summary, the submillimeter cores with concentrations higher than
0.6 are all star forming with a decreasing trend toward lower
concentrations, but it is worth noting that a number of low
concentration/flux cores do have embedded YSOs, so this distinction is
not unique. In \S\ref{sample} we use these results together with the
results concerning the spatial distribution of embedded YSOs and their
colors from the above discussions to establish an unbiased list of
embedded YSOs.  
\section{Comparison to the overall cloud environment}\label{overallcloud}
\subsection{Extinction as a tracer}\label{extincttracer}
\begin{figure}
\plotone{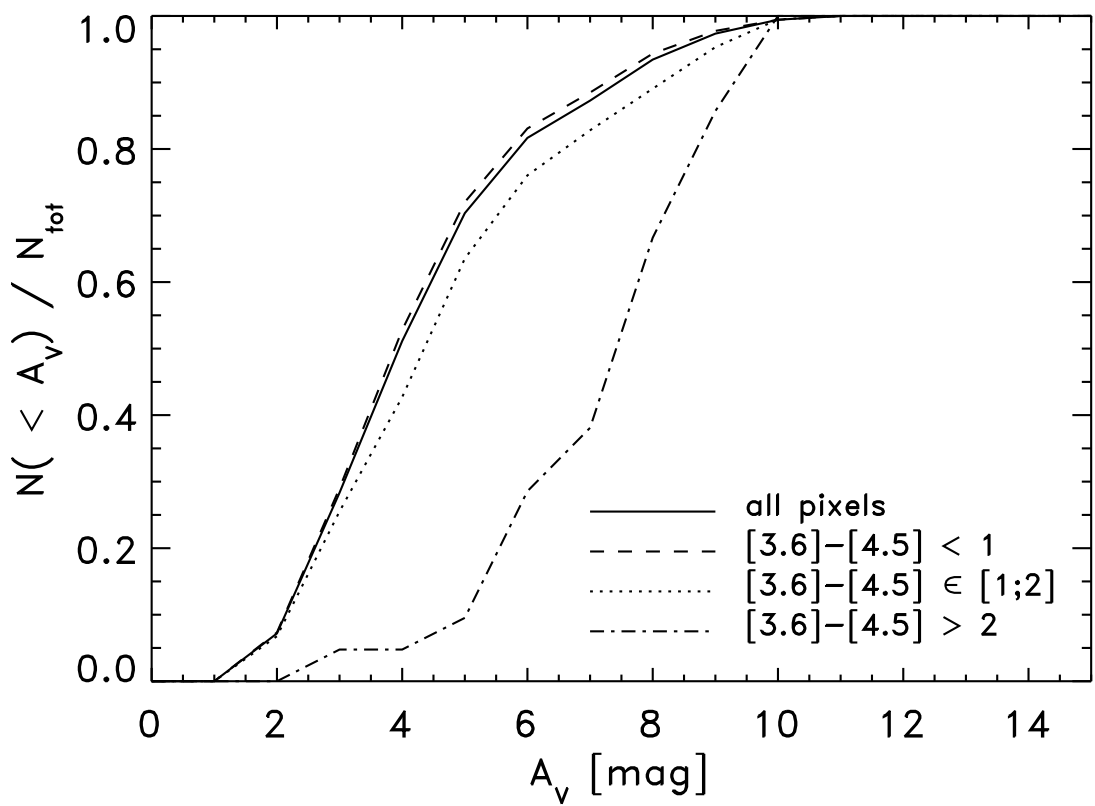}
\caption{Cumulative fraction of mid-infrared sources with differing
  $[3.6]-[4.5]$ colors as a function of $A_V$ from the extinction
  maps. The four different lines indicate all the pixels from the
  extinction map covered by the mid-infrared observations (black
  solid), and mid-infrared sources with $[3.6]-[4.5] < 1$ (dashed), $1
  \le [3.6]-[4.5] \le 2$ (dotted) and $[3.6]-[4.5] > 2$
  (dashed-dotted). \label{extinct_cumm}}
\end{figure}

The submillimeter cores are in general found in regions of high
extinction \citep{enoch06,kirk06} suggesting an extinction threshold
for the formation of the dust cores \citep[see also][]{johnstone04}. A
similar trend is seen when comparing the location of the mid-infrared
sources to the extinction maps from COMPLETE \citep{alves06,kirk06}:
Fig.~\ref{extinct_cumm} shows the cumulative distribution of red MIPS
sources as a function of $A_V$. From this figure it is found that 90\%
of the YSOs with $[3.6]-[4.5] > 2$ are at $A_V > 5$, which is similar
to the extinction threshold for the cores in Perseus
\citep{enoch06,kirk06}, as one would expect from the relation between
the dense cores and the embedded YSOs. This of course also begs the
question: are the observed trends in colors simply reflecting the
overall cloud extinction toward the low-mass protostars? It is
unlikely since even a fairly high amount of extinction only produces a
small shift in $[3.6]-[4.5]$ colors: according to the extinction
measurements of \cite{huard06} for an $A_V = 50$ $(A_K = 5.6)$ a shift
of $E_{[3.6]-[4.5]}=0.36$ is expected. This is also similar to the
conclusion of studies examining the properties of young stellar
objects in color-color diagrams \citep[e.g.,][]{allen04} which
generally find that the young stellar objects have infrared excesses
that cannot be explained by the extinction from the cloud material,
but rather are due to a combination of the extinction in the inner
parts of the protostellar envelopes and emission from warm dust in the
envelope and disk.  
\subsection{Submillimeter flux as a tracer}\label{submmtracer}
\begin{figure}
\plotone{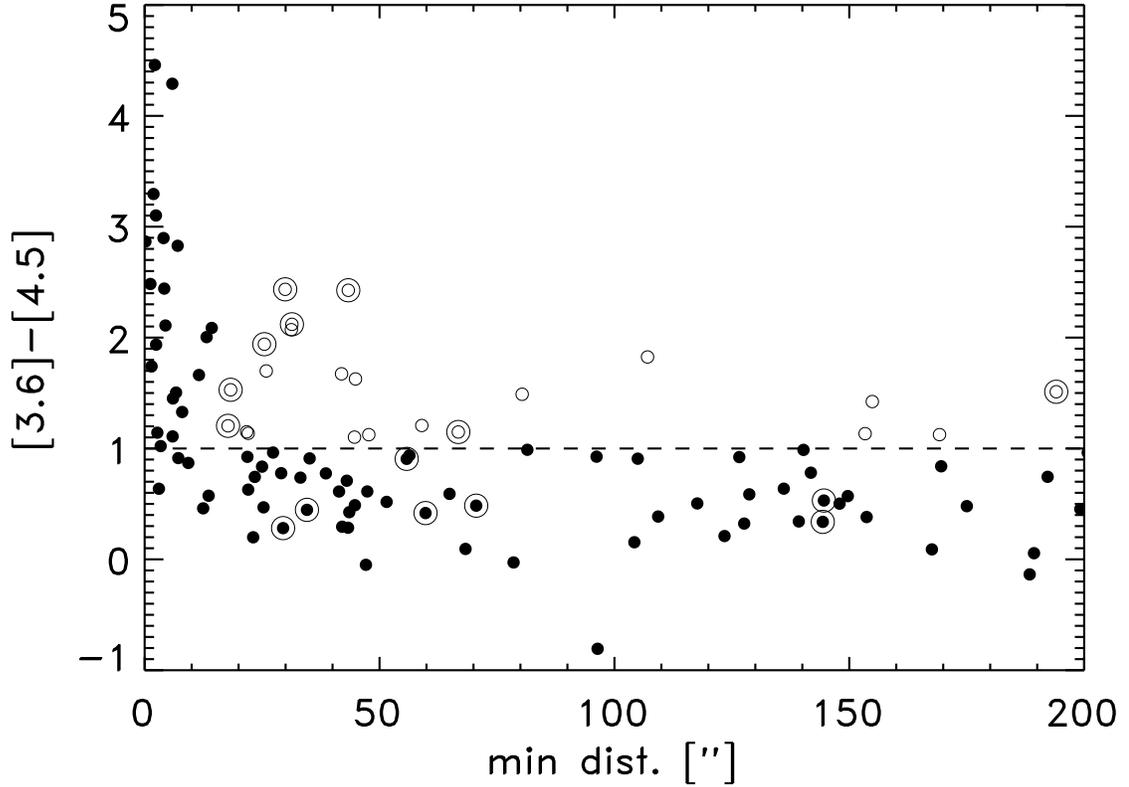}
\caption{$[3.6]-[4.5]$ color of each MIPS source vs. its distance to
  the nearest submillimeter core. Sources with distances larger than
  15$''$ to their nearest submillimeter core and $[3.6]-[4.5] > 1.0$
  have been indicated by open circles. A second larger circle around
  any symbol (filled or open circle) indicates a mid-infrared source
  with a distance to the nearest submillimeter core larger than 15$''$
  and with a $[8.0]-[24]$ color greater than 4.5 (those sources are
  also indicated by open circles in
  Fig.~\ref{inverse_45}).}\label{inverse_12}
\end{figure}

\begin{figure}
\plotone{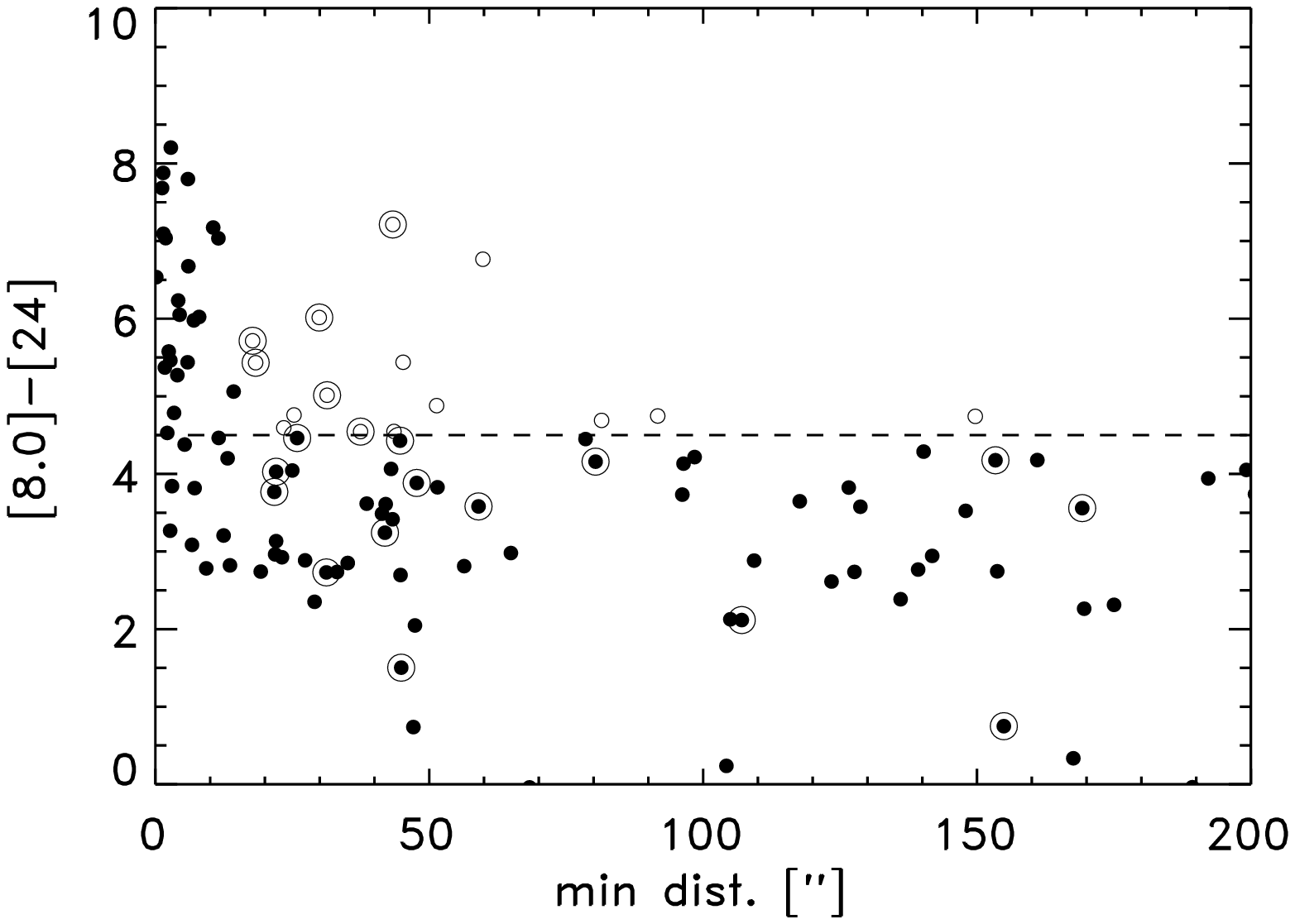}
\caption{$[8.0]-[24]$ color of each MIPS source vs. its distance to
  the nearest submillimeter core. Sources with distances larger than
  15$''$ to their nearest submillimeter core and $[8.0]-[24] > 4.5$
  have been indicated by open circles. A second larger circle around
  any symbol (filled or open circle) indicates a mid-infrared source
  with a distance to the nearest submillimeter core larger than 15$''$
  and with a $[3.6]-[4.5]$ color greater than 1.0 (those sources are
  also indicated by open circles in
  Fig.~\ref{inverse_12}).}\label{inverse_45}
\end{figure}

\begin{figure}
\plotone{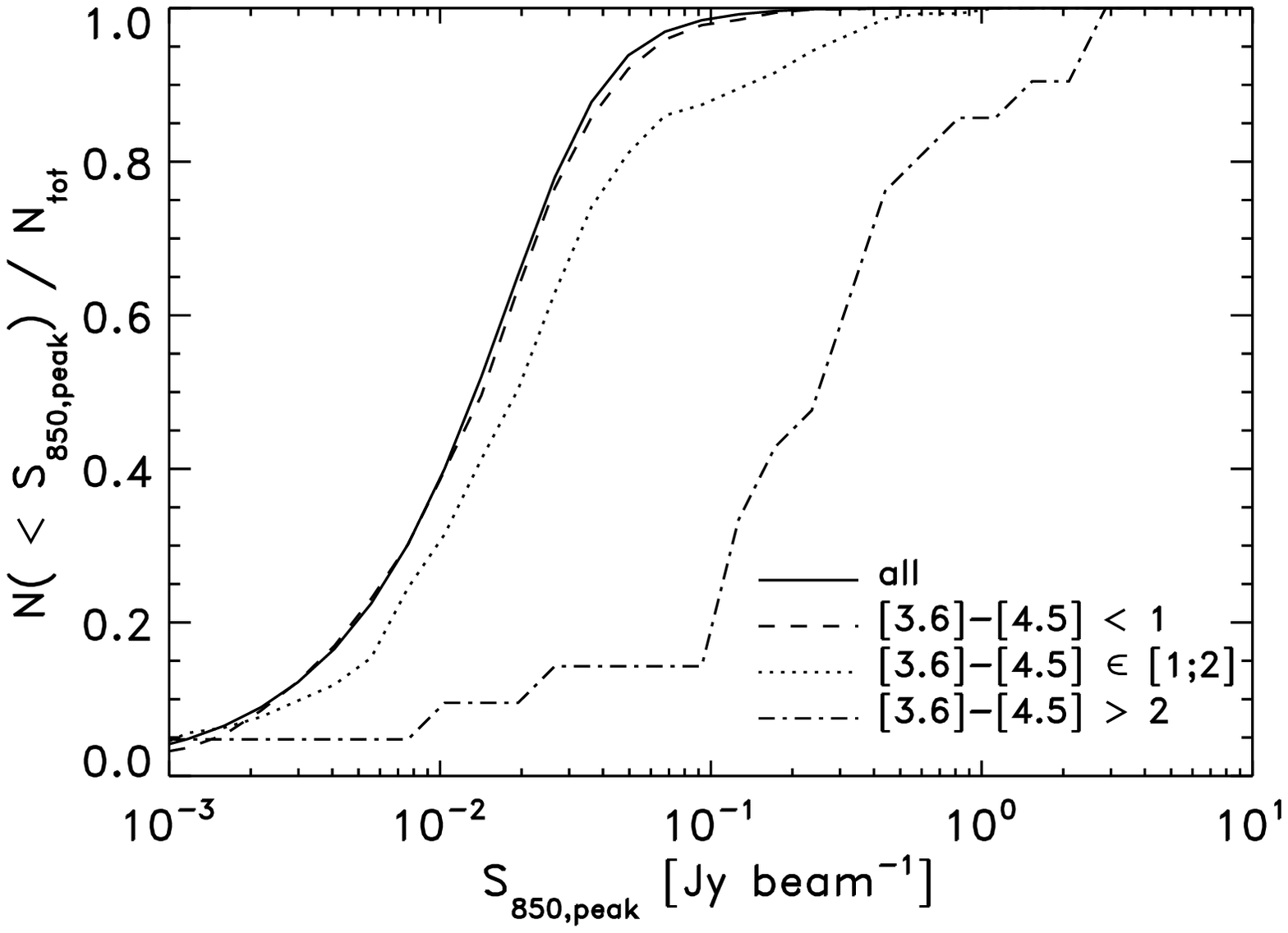}
\caption{Cumulative number of mid-infrared sources (with S/N of 7 or
  greater in IRAC bands 1 and 2) as a function of SCUBA flux at the
  position of the MIPS source; divided according to $[3.6]-[4.5]$
  colors as in Fig.~\ref{extinct_cumm}.}\label{cumu}
\end{figure}

The discussion in \S\ref{cluster} was based on the question ``what is
the nearest MIPS source to any given SCUBA core?''. To fully address
the completeness of this method we also have to ask the inverse
question: ``what is the distance for any given red source to its
nearest submillimeter core?''. Fig.~\ref{inverse_12} and
\ref{inverse_45} show the $[3.6]-[4.5]$ and $[5.8]-[8.0]$ colors of
all the MIPS sources as a function of distance to the nearest SCUBA
core. In both Fig.~\ref{inverse_12} and \ref{inverse_45}, a number of
red sources are found with distances of 20--100\arcsec\ to their
nearest cores. This does not mean that these red objects are not
associated with high levels of extinction. Fig.~\ref{cumu} shows the
cumulative distribution of SCUBA fluxes at the positions of YSOs with
$[3.6]-[4.5]$ colors lower than 1, between 1 and 2 and higher than
2. The reddest objects ($[3.6]-[4.5] > 2$) are found in regions with
SCUBA fluxes higher than 0.1~Jy~beam$^{-1}$ (90\% of the red sources),
corresponding to an extinction $A_V \gtrsim 7$ for a dust temperature
of 15~K and an 850~$\mu$m dust opacity per unit total (dust+gas) mass
of 0.0182~cm$^{2}$~g$^{-1}$ \citep{ossenkopf94}. In effect, the amount
of extinction will be lower for protostars which are heating their
envelopes to higher temperatures, but on the other hand, since high
amounts of extinction are required to explain the very red colors, the
actual central extinction is significantly higher than that calculated
on basis of the SCUBA flux due to beam dilution. It should also be
contrasted to the average extinction from the extinction maps
(\S\ref{extincttracer}) which measures the larger scale structure of
the cloud.  

Visual inspection of the SCUBA data at the positions of the MIPS
sources with distances to their nearest core larger than 15$''$,
$[3.6]-[4.5] > 1.0$, and $[8.0]-[24] > 4.5$ reveals that almost all of
these are associated with dust continuum peaks with low fluxes. This
demonstrates that in the confused regions, in particular, the exact
identification of cores is somewhat dependent on the algorithm,
assumed noise level etc. as demonstrated by a comparison between the
lists of cores in Perseus from \cite{kirk06} and \cite{hatchell05}
(both based on similar, but not identical, SCUBA 850~$\mu$m data, and
both processed differently) and that of \cite{enoch06} based on
Bolocam 1.1~mm data \citep[see also discussion in][]{young06}. Still,
by imposing the color criteria above, it is possible to pick out the
remaining YSO candidates which are missed due to the lack of a core
extracted from the submillimeter maps for those reasons.

\section{Discussion}\label{discuss}
\subsection{Constructing an unbiased sample of embedded YSOs}\label{sample}
Based on the above discussions it is possible to create a relatively
unbiased list of 49 deeply embedded YSOs in Perseus
(Table~\ref{finallist}) by selecting sources that fulfill any one of
the following three criteria:
\begin{description}
\item[A.] MIPS catalog sources with $[3.6]-[4.5] > 1$ and
  $[8.0]-[24] > 4.5$ \emph{(23 sources)}.
\item[B.] MIPS catalog sources with distances less than 15\arcsec\ to
  their nearest core \emph{(39 sources)}.
\item[C.] Submillimeter cores with concentrations higher than 0.6  \emph{(21 sources)}.
\end{description}
The first criterion will select all MIPS sources that are embedded
YSOs, but miss sources that are not detected in IRAC bands 1 or 2 -
either due to confusion with outflows or simply because they in fact
are very deeply embedded. That list will also include sources which
are not necessarily directly associated with a submillimeter core due
to confusion in the submillimeter map, poorly defined cores, or simply
the sensitivity of the submillimeter observations. The second
criterion will select the most deeply embedded YSOs with steeply
increasing SEDs like IRAS~16293-2422 \citep{iras16293letter}. As
pointed out in \S\ref{midir} most of the sources selected by these
first two criteria will be overlapping but will miss sources
that are saturated at 24~$\mu$m. Such sources will be selected by the
third criterion, which does not include all the mid-infrared sources
under {\bf A} and {\bf B} as shown in Fig.~\ref{conc_histo}. Neither
the mid-infrared source lists nor the submillimeter core
identifications by themselves can be used unambiguously to define a
sample of embedded objects. Applying for example only criterion {\bf
  A} which is based solely on the mid-infrared colors or criterion
{\bf C} which is based solely on the submillimeter maps would pick out
only 40--50\% of the sources, but used in conjunction they will form a
relatively unbiased list of YSOs complete down to the resolution of
the mid-infrared data. It should also be emphasized that these YSOs
constitute an important addition of embedded sources to the list of
YSOs discussed in \cite{perspitz} - although they only constitute a
small fraction ($\approx$~10\%) of the overall YSO population in the
cloud.  
\begin{table*}
\caption{List of embedded YSOs in Perseus.}\label{finallist}
{\tiny \begin{tabular}{ccccccccl}\hline\hline
         & \multicolumn{2}{c}{Position\tablenotemark{a}} &               &              &               &               &            &  \\
Number   & RA          & DEC            & \multicolumn{2}{c}{Mid-IR Colors} & Separation\tablenotemark{b} & Conc.\tablenotemark{c} & Code\tablenotemark{d} & Other identifiers\tablenotemark{e} \\
         & (J2000)     &  (J2000)       & $[3.6]-[4.5]$ & $[8.0]-[24]$ &      [$''$]   &               &            &                  \\\hline
       1 & 03 25 22.36 & +30 45 13.6 &  2.9 &  6.5 &  0.2 & 0.70 & ABC & L1448-IRS2 / IRAS 03222+3034             \\
       2 & 03 25 36.48 & +30 45 23.2 &  2.8 &  6.0 &  7.0 & 0.87 & ABC* & L1448-N(A)                               \\
       3 & 03 25 38.87 & +30 44 06.0 &  1.7 &  7.1 &  1.5 & 0.83 & ABC & L1448-C(N)                               \\
       4 & 03 26 37.46 & +30 15 28.2 &  2.4 &  6.2 &  4.2 & 0.39 & AB$-$ &                                          \\
       5 & 03 27 38.27 & +30 13 58.5 &  0.6 &  3.8 &  3.0 & 0.31 & $-$B$-$* & [TPB97] L1455-FIR2                       \\
       6 & 03 27 39.11 & +30 13 02.8 &  3.3 &  7.0 &  1.9 & 0.59 & AB$-$* & L1455-IRS1 / IRAS 03245+3002             \\
       7 & 03 27 43.25 & +30 12 28.9 &  2.1 &  6.1 &  4.4 & 0.57 & AB$-$ & L1455-IRS4                               \\
       8 & 03 27 47.69 & +30 12 04.4 &  0.9 &  2.8 &  9.3 & 0.48 & $-$B$-$* &                                          \\
       9 & 03 28 32.55 & +31 11 04.8 & -0.4 &  5.4 &  1.7 & 0.27 & $-$B$-$ &                                          \\
      10 & 03 28 34.53 & +31 07 05.5 &  1.5 &  3.1 &  6.7 & 0.30 & $-$B$-$* &                                          \\
      11 & 03 28 37.11 & +31 13 28.3 &  2.8 &  ... &  ... & 0.66 & $-$$-$C & IRAS 03255+3103                          \\
      12 & 03 28 39.11 & +31 06 01.6 &  1.3 &  6.0 &  8.0 & 0.32 & AB$-$ & (ass. HH 340)                            \\
      13 & 03 28 40.62 & +31 17 56.5 &  ... &  4.4 &  5.3 & 0.51 & $-$B$-$ &                                          \\
      14 & 03 28 45.31 & +31 05 41.9 &  2.4 &  7.2 & 43.3 & ...  & A$-$$-$ & IRAS 03256+3055                          \\
      15 & 03 28 55.59 & +31 14 37.5 &  ... &  ... &  ... & 0.87 & $-$$-$C & NGC1333-IRAS2A                           \\
      16 & 03 28 57.36 & +31 14 15.9 &  2.1 &  5.0 & 31.4 & ...  & A$-$$-$* & NGC1333-IRSA2B                           \\
      17 & 03 28 59.55 & +31 21 46.7 &  0.6 &  2.8 & 13.6 & 0.67 & $-$BC* &                                          \\
      18 & 03 29 00.61 & +31 12 00.4 &  ... &  ... &  2.4 & 0.48 & $-$B$-$ &                                          \\
      19 & 03 29 01.66 & +31 20 28.5 &  ... &  ... &  ... & 0.67 & $-$$-$C & (ass. SVS12/ASR114)      \\
      20 & 03 29 03.30 & +31 15 55.5 &  ... &  ... &  ... & 0.77 & $-$$-$C & NGC1333-SVS13                         \\
      21 & 03 29 04.09 & +31 14 46.6 &  1.5 &  6.7 &  6.0 & 0.53 & AB$-$ & HH7-11 MMS6                              \\
      22 & 03 29 10.53 & +31 13 30.7 &  ... &  ... &  0.8 & 0.86 & $-$BC & NGC1333-IRAS4A                           \\
      23 & 03 29 10.72 & +31 18 20.5 &  ... &  7.2 & 10.6 & 0.74 & $-$BC2 &                                          \\
      24 & 03 29 11.29 & +31 18 31.3 &  ... &  8.2 &  2.8 & 0.74 & $-$BC2 &                                          \\
      25 & 03 29 12.07 & +31 13 01.8 &  4.3 &  5.4 &  5.9 & 0.86 & ABC & NGC1333-IRAS4B\tablenotemark{f}\\
      26 & 03 29 13.62 & +31 13 57.9 &  3.1 &  ... &  2.4 & 0.58 & $-$B$-$ & NGC1333-IRAS4C                           \\
      27 & 03 29 17.21 & +31 27 46.2 &  1.1 &  7.8 &  5.9 & 0.54 & AB$-$ & (ass. [LMG94] Per 4)                     \\
      28 & 03 29 18.25 & +31 23 19.9 &  1.7 &  4.5 & 11.6 & 0.55 & $-$B$-$*2 & (ass. HH335)                             \\
      29 & 03 29 18.73 & +31 23 25.4 &  0.5 &  3.2 & 12.5 & 0.55 & $-$B$-$*2 &                                          \\
      30 & 03 29 23.50 & +31 33 29.4 &  ... &  7.0 & 11.5 & 0.43 & $-$B$-$ & IRAS 03262+3123                          \\
      31 & 03 29 51.89 & +31 39 05.6 &  ... &  7.9 &  1.4 & 0.53 & $-$B$-$ & IRAS 03267+3128 / ([LMG94] Per 5)   \\
      32 & 03 31 21.01 & +30 45 30.0 &  1.9 &  5.6 &  2.4 & 0.71 & ABC & IRAS 03282+3035                          \\
      33 & 03 32 18.03 & +30 49 46.9 &  1.0 &  4.8 &  3.4 & 0.82 & ABC & IRAS 03292+3039                          \\
      34 & 03 33 13.81 & +31 20 05.2 &  2.1 &  5.1 & 14.3 & 0.31 & AB$-$ & (ass. [LMG94] Per 9B)                    \\
      35 & 03 33 14.41 & +31 07 10.8 &  2.4 &  6.0 & 29.9 & ...  & A$-$$-$*2 & B1-SMM3                                  \\
      36 & 03 33 16.49 & +31 06 52.3 &  ... &  ... &  2.7 & 0.55 & $-$B$-$2 & B1-d                                     \\
      37 & 03 33 16.67 & +31 07 55.1 &  2.9 &  5.3 &  4.0 & 0.38 & AB$-$ & B1-a / IRAS 03301+3057                   \\
      38 & 03 33 17.87 & +31 09 31.8 &  4.5 &  4.5 &  2.2 & 0.84 & ABC* & B1-c                                     \\
      39 & 03 33 20.34 & +31 07 21.4 &  2.0 &  4.2 & 13.2 & 0.74 & $-$BC* & B1-b                                     \\
      40 & 03 33 27.31 & +31 07 10.2 &  2.5 &  7.7 &  1.2 & 0.32 & AB$-$ & (ass. HH789)                             \\
      41 & 03 43 50.99 & +32 03 24.7 &  1.3 &  5.5 &  2.7 & 0.65 & ABC &                                          \\
      42 & 03 43 51.03 & +32 03 08.0 &  1.2 &  5.7 & 17.8 & ...  & A$-$$-$* &                                          \\
      43 & 03 43 56.91 & +32 03 04.2 &  ... &  ... &  3.1 & 0.79 & $-$BC & IC348-MMS                                \\
      44 & 03 43 57.32 & +32 00 47.6 &  ... &  ... &  8.3 & 0.78 & $-$BC2 & HH211-FIR (?)                            \\
      45 & 03 43 57.64 & +32 00 44.8 &  ... &  ... & 13.1 & 0.78 & $-$BC2 & (ass. HH211?)                             \\
      46 & 03 43 59.41 & +32 00 35.5 &  2.6 &  4.5 & 37.5 & ...  & A$-$$-$ & (ass. HH211?)                           \\
      47 & 03 44 02.40 & +32 02 04.7 &  1.5 &  5.4 & 18.3 & ...  & A$-$$-$ &                                          \\
      48 & 03 44 43.32 & +32 01 31.6 &  0.9 &  3.8 &  7.2 & 0.62 & $-$BC* & IRAS 03415+3152                          \\
      49 & 03 47 41.61 & +32 51 43.9 &  1.1 &  3.3 &  2.7 & 0.59 & $-$B$-$* & B5-IRS1 / IRAS 03445+3242                \\ \hline
\end{tabular}}

\tablenotetext{a}{Coordinates given in format of (hh mm ss.ss) for right ascension and (dd mm ss.s) for declination. Coordinates for sources selected on basis of the core concentration only (code ``C'') refer to position of SCUBA core.}
\tablenotetext{b}{Separation between the mid-infrared source and the nearest submillimeter core.}
\tablenotetext{c}{Concentration of submillimeter core from \cite{kirk06}.}
\tablenotetext{d}{Code for selection of source according to which of the criteria (A, B, or C) in \S\ref{sample} applies for the given source. Sources included in the list of \cite{perspitz} are furthermore identified by ``*''. Cores associated with multiple YSOs (within 15\arcsec) in Table~\ref{multiplicity} is furthermore identified by a ``2'' in this code.}
\tablenotetext{e}{Refers to main designation in the SIMBAD database, commonly used identifier or association (marked ``ass.''). Specific references: [TPB97]: \cite{tapia97}, [LMG94]: \cite{ladd94}.} \tablenotetext{f}{c2d catalog refers to position of bright shock observed in the IRAC bands.}
\end{table*}
\subsection{Colors of deeply embedded protostars}\label{evolution}
It is interesting to examine whether there is any evolution of the
colors of the embedded YSOs compared to the properties of their
parental cores. For example do the more massive cores generally have
more deeply embedded protostars with redder mid-infrared colors? As
shown in Fig.~\ref{flux_histo}, cores with stronger submillimeter
fluxes all have embedded sources but a significant number of
``fainter'' cores also appear to be forming stars.

\begin{figure}
\resizebox{\hsize}{!}{\includegraphics{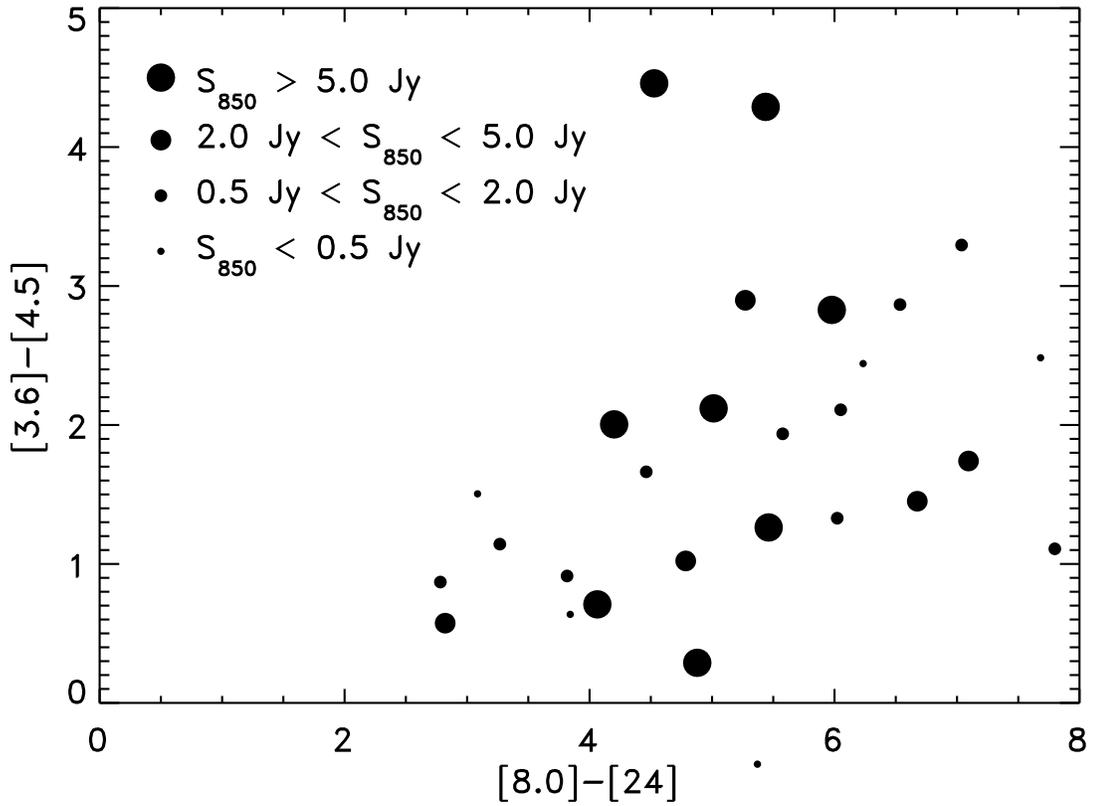}}
\caption{$[3.6]-[4.5]$ vs. $[8.0]-[24]$ color-color diagrams for the
  embedded YSOs. For each YSO the size of the symbol indicates the
  total flux of the parental core from the submillimeter observations
  \citep{kirk06}.}\label{evol_diagram}
\end{figure}

Fig.~\ref{evol_diagram} shows the $[3.6]-[4.5]$ and $[8.0]-[24]$
colors of the embedded YSOs with the fluxes of their parental cores
indicated. There are two interesting effects to note: (i) this diagram
includes all the MIPS sources located within 15\arcsec\ of a
submillimeter core and for those sources there is a clear trend across
the diagram with an increase in both colors (from zero; or no infrared
excess) toward higher values, possibly approaching a maximum value of
their $[8.0]-[24]$ colors of about 6--8. (ii) If we apply our
criterion of $[3.6]-[4.5] > 1$ for embeddedness the $[8.0]-[24]$
colors do not actually increase systematically with flux, but are
rather scattered around a value of about 6. Some of the brighter SCUBA
cores have YSOs with very red colors, but note also that some
mid-infrared sources associated with fairly bright cores have
$[3.6]-[4.5] \lesssim 1.0$.

The upturn in colors is roughly as predicted by the models of
\cite{whitney03b} but this diagram extends to much bluer [8.0]-[24]
colors than predicted in their models. On the other hand the
$[3.6]-[4.5]$ colors are much redder than predicted from the models.
\citeauthor{whitney03b} find that only a few deeply embedded YSOs
should be detectable at the shortest IRAC wavelengths. That we
actually detect a significant fraction of the deeply embedded
protostars might again reflect that the envelopes have larger cleared
out cavities similar to the case of IRAS~16293-2422A
\citep{iras16293letter}, in contrast to the assumption of the models
of \cite{whitney03b} which extend into 7 stellar radii. At such small
inner radii, the mid-infrared emission from the central protostar
would be completely absorbed by the surrounding envelope and the
object not detectable at the IRAC wavelengths. The fact that most of
these sources are detectable at 3.6 and 4.5~$\mu$m suggest that this
is not the case and $[3.6]-[4.5]$ color therefore is a potentially
interesting diagnostic of the properties of the disks and inner
envelopes of deeply embedded protostars.  
\subsection{Constraints on star formation}
As pointed out in the introduction many important constraints on
low-mass star formation hinge on statistical arguments such as using
the relative number of pre- and protostellar cores to estimate their
lifetimes. Previous studies have often been based on compilations of
different surveys not necessarily observed under similar conditions or
with the same techniques and have often required the combination of
data for many clouds to obtain statistically meaningful samples. The
strength of the combination of the Spitzer and SCUBA data presented in
this paper is that it represents a relatively uniform survey of just
one cloud and still has enough cores and protostars that strong
statements can be made without many of the caveats introduced
otherwise. We here discuss the constraints on three important aspects
of low-mass star formation introduced by this study.

\subsubsection{Motions and dispersal of protostars}\label{dispersal}
One of the debated issues in current theories of star formation is the
importance of the motions of embedded YSOs through the medium on their
accretion rates, i.e., basically whether Bondi-Hoyle accretion
\citep{bondi44} applies. The simulations of
\cite{bonnell97,bonnell01a} and \cite{bate03} for example show that
the motions of the embedded YSOs through their ambient medium will
have a significant impact on their accretion histories and thereby the
resulting distribution of YSOs as a function of mass. On the other
hand based on a study of locations of YSOs in different evolutionary
stages from Taurus, \cite{hartmann02} argues that the velocity
dispersion of the YSOs once formed is small $\lesssim
0.2$~km~s$^{-1}$. The narrowness of the distribution of separations
between the mid-infrared sources and submillimeter cores
(Fig.~\ref{clustdia_hist}) indicates that the dispersal of newly
formed stars in Perseus is similar to that in Taurus.

As shown in Fig.~\ref{clustdia_hist} the typical separations between
the mid-infrared sources and submillimeter cores are less than
$l\approx$~10\arcsec. Assuming a typical lifetime of the deeply
embedded protostars of $t=10^5$~years, the velocity dispersion is:
\begin{equation}
\delta v=l/t=0.1\,{\rm km~s}^{-1}\, \left(\frac{l}{10''}\right)\, \left(\frac{d}{250\, {\rm pc}}\right)\,\left(\frac{t}{10^5\, {\rm years}}\right)^{-1}
\end{equation}
which is smaller than the isothermal sound speed,
$c_s=\sqrt{\frac{kT}{m}} \approx 0.2$~km~s$^{-1}$ for a molecular
cloud with a temperature of 10--15~K. \cite{walsh04} examined the
motions of dense cores traced by N$_2$H$^+$ $J=1-0$ emission with
respect to less dense ambient envelopes traced by $^{13}$CO and
C$^{18}$O $J=1-0$. They found only small velocity differences between
the ``cores'' and ``envelopes'', which they took as evidence that
dense cores did not have significant velocities with respect to their
surroundings. \cite{ayliffe06} examined the models of \cite{bate03}
using a similar approach as that in \cite{walsh04} and also found
small velocities of the cores relative to their lower density ambient
envelopes. It therefore appears that those simulations are not
invalidated by the observations of \citeauthor{walsh04} What is shown
by this analysis, however, is that the newly formed stars still
embedded within their parental cores also do not have highly
supersonic velocities and therefore do not seem to travel a
substantial distance during their main accretion phase. It is also
interesting to note the similarity to Taurus: even though Taurus is
thought to be forming low-mass stars in a more quiescent fashion
compared to Perseus, the velocity dispersions are very similar and
significantly below the 1~km~s$^{-1}$ virial speeds inferred in models
of protostars accreting while moving through the molecular cloud (and
halting accretion when ``leaving'' the densest regions).

\subsubsection{Time scales}\label{timescales}
This study finds that approximately half the SCUBA cores from the
study of \cite{kirk06} have embedded YSOs. This suggests that the
dissipation time scale of the cores once a protostar has been formed
is similar to the time scale over which the cores evolve through the
part of the prestellar stage observable by the SCUBA
studies. \cite{lee99} studied a sample of optically identified cores
and found that only 24\% had embedded YSOs, which would imply a time
scale for their optically identified, starless cores three times
longer than the duration of the protostellar stage. A comparison
between the numbers from this study and those of \citeauthor{lee99}
therefore suggests that the SCUBA maps only detect the denser stages
in the evolution of prestellar cores whereas the sample of optically
identified cores by \citeauthor{lee99} also includes less evolved
cores. In a study of more dense NH$_3$ cores, \cite{jijina99} found
that the ratio of starless to stellar cores were close to unity or
lower in a wide range of environments, likewise implying similar time
scales for the evolution through the pre- and protostellar
stages. \cite{kirk05} observed a large sample of 52 starless NH$_3$
cores with SCUBA and found that the cores which were bright at
submillimeter wavelengths (i.e., having peak 850~$\mu$m flux densities
larger than 170~mJy~beam$^{-1}$) constituted approximately half of the
cores detected by SCUBA - and that the cores detected with SCUBA
constituted approximately half of all the studied NH$_3$
cores. Comparison across such different samples should naturally be
treated with care dealing with systematic uncertainties, selection
effects, etc. The big strength of the sample forming the basis of this
paper is again that the sources are selected from one large scale
systematic and relatively unbiased set of observations with uniform
sensitivity.

Assuming a typical lifetime of the deeply embedded stages studied here
of $1\times 10^5$~years,\footnote{\cite{wardthompsonppv} quote a best
  estimate of the time scale of the evolution of YSOs through the
  embedded phases of $2\pm 1\times 10^5$~years from previous studies -
  which includes parts of the ``less embedded'' Class I stages where
  the cores associated with the YSOs are not picked up by the SCUBA
  surveys.} the total duration of the starless core stage would be
$3\times 10^5$~years - with the SCUBA maps revealing the last $1\times
10^5$~years of this evolution where the central density increases
above $5\times 10^4-1\times 10^5$~cm$^{-3}$ (J.~\citealt{kirk05},
H.~\citealt{kirk06}). As pointed out by \citeauthor{lee99} these time
scales are significantly shorter than those for typical ambipolar
diffusion models unless the cores are marginally subcritical
\citep{ciolek01}. They are similar to the time scale for depletion in
protostellar cores of $10^{5\pm 0.5}$~years derived by
\cite{coevollet} based on the measurements of chemical profiles of
pre- and protostellar cores - again suggesting that the SCUBA dust
condensations reflect the stage of prestellar cores where the density
is high and significant depletion can start to occur.

We can naturally also compare these numbers to the time scales for the
later stages of protostellar evolution. As pointed out above the
objects identified here represent YSOs in their first $10^{5}$~years
of evolution and are an important addition to the sample of
\cite{perspitz}. That study was missing some of these deeply embedded
objects but more complete in terms of evolved Class~I and Class~II
objects which evolve over time scales of approximately 10$^6$~years. The
number of sources in this list compared to those of \cite{perspitz}
appears to be consistent with the ratio between those time scales of
about 10:1 - or alternatively if we accept those time scales as
``known'': that star formation in Perseus has occurred at a fairly
constant rate continuing to current times.

\subsubsection{Star formation efficiency}\label{SFE}
We can take the arguments a step further and estimate the star
formation efficiency of the submillimeter cores and the entire cloud
complex. Given the amount of mass in each core from the submillimeter
observations the star formation efficiency can be estimated:
\begin{equation}
{\rm SFE}=\frac{M_{\rm YSO}}{M_{\rm YSO}+M_{\rm core}}
\end{equation}
where $M_{\rm YSO}$ is the mass of the new formed stars and $M_{\rm
  core}$ the mass of the cores. The total mass of the 72 SCUBA cores
in the Perseus cloud was found to be $M_{\rm core}=$106~$M_\odot$
\citep{kirk06} and counting only those YSOs within 15$''$ of one of
these submillimeter cores we find 43 YSOs. Assuming that the masses of
the newly formed stars are 0.3--0.5~$M_\odot$, the star formation
efficiency for the entire Perseus cloud complex is estimated to be
$\approx$~10--15\%. This is comparable to the star formation
efficiency of nearby embedded clusters \citep{lada03}.  Approximately
half of the total core mass in the cloud resides in the cores in
NGC~1333, but doing the same calculation as above the star formation
efficiency for just those cores is found to be similar to the value
for the entire cloud. This again implies that significant star
formation is currently occurring in the cores ``outside'' the main
NGC~1333 and IC~348 clusters -- in the dense regions associated with
L1448, L1455, Barnard~1 and Barnard~5 amongst others -- similar to the
conclusion of \cite{perspitz}. It should be noted that the mass
estimates are sensitive to the assumptions about the dust temperature
and the dust opacities and to the observational techniques being more
or less sensitive to extended structure, for example.

The star formation efficiency quoted above is an estimate of the
current core star formation efficiency and does not represent the
actual efficiency of assembling the cores from the overall cloud
environment: As pointed out by \cite{kirk06} (see also
\cite{johnstone04} for a similar result for the Ophiuchus molecular
cloud complex) only a small fraction ($<$~0.5\%) of the total cloud
mass resides in these dense cores with the main fraction ($>$~85\%) of
the cloud mass from the extinction maps ($\approx$~18,500~$M_\odot$;
\cite{kirk06}) distributed at low $A_V < 5$. That number cannot be
compared to the current mass of deeply embedded protostars from this
analysis: a high number of young stellar objects have already formed
in the Perseus molecular cloud with for example \cite{perspitz}
identifying a ``high quality'' sample of 400 YSOs of Class II or
earlier. These Spitzer observations are furthermore not picking out a
significant number of more evolved, i.e., Class III YSOs with smaller
amount of infrared excesses. With these higher number of YSOs
distributed over all evolutionary stages and the total cloud mass, the
time averaged star formation efficiency over the entire evolution of
the cloud until present time is probably closer to a few percent,
which includes the efficiency of assembling cores from the overall
cloud dust mass in the first place and subsequently the efficiency of
these cores actually forming stars. In this context it should also be
emphasized that the observed mass distribution of SCUBA core masses is
a current snapshot of potentially star forming cores. With their
lifetimes of about $10^5$~years (Sect.~\ref{timescales}) it is not
unreasonable to expect that at least 10 times as many cores have
previously existed in the cloud and formed what are currently observed
as Class II and III young stellar objects.

\section{Conclusions}
This paper has presented a comparison between the dust condensations
from SCUBA maps and mid-infrared sources from Spitzer
observations. This study represents a large systematic census of the
embedded YSO population and thereby allows statements from a large
sample of sources. The main conclusions of this paper are:

\begin{enumerate}
\item The mid-infrared sources with 24~$\mu$m detections and red
  colors are found to be located close to the center of the SCUBA
  cores mapped by \cite{kirk06} - typically within 15$''$ (less than
  half a core radius) of their peaks. The sources are found to have
  characteristic red colors with $[3.6]-[4.5] > 1.0$ and $[8.0]-[24] >
  4.5$ but do not show similarly red $[5.8]-[8.0]$ colors. The most
  deeply embedded objects are found in regions with high extinction,
  $A_V \ge 5$ (90\% of the objects with $[3.6]-[4.5] > 2$), similar to
  the extinction threshold for the SCUBA cores. Still, the red colors
  are not purely an effect of cloud extinction.
\item All the SCUBA cores with high ``concentrations'' have embedded
  YSOs, but also a non-negligible number of cores with low
  concentrations have embedded YSOs making this distinction
  ambiguous. A few of the strongest sources in NGC~1333 are not picked
  up by the above criteria described under point 1. - due to
  saturation and confusion. The cores associated with those sources
  can be picked out by their high concentrations and submillimeter
  fluxes.
\item From the above considerations a relatively unbiased sample of
  deeply embedded YSOs can be constructed. In this fashion we identify
  49 embedded YSOs associated with SCUBA cores from the list of
  \cite{kirk06}. Out of 72 SCUBA cores 40 are found to have associated
  MIPS sources (including four cores which associated with saturated
  MIPS sources not included in the c2d catalog). Only 3 SCUBA cores
  are found to have more than one MIPS sources within 15$''$.
\item The narrowness of the spatial distribution of red sources within
  the SCUBA cores suggests that the velocities of the newly formed
  protostars relative to the cores are low - close to the isothermal
  sound speed. This argues against the suggestion that the motions of
  protostars relative to the ambient cloud regulates the time scale
  over which significant accretion can occur.
\item From the number of SCUBA cores with and without embedded YSOs
  (40 and 32, respectively) the time scale of the evolution through
  the dense prestellar stages, where the cores are recognized in the
  submillimeter maps and their central densities are larger than
  $5\times 10^4-1\times 10^5$~cm$^{-3}$, is similar to the time scale
  for evolution through the embedded protostellar stages. This result
  further implies that the observable SCUBA core stage constitutes a
  third of the lifetime of less dense prestellar cores, e.g., those
  picked up in optical or NH$_3$ surveys.
\item The current star formation efficiency estimated from the mass of
  the embedded protostars and the SCUBA cores is found to be
  $\approx$~10--15\%, similar to the star formation efficiency of
  nearby embedded clusters. If we also take into account the
  efficiency of assembling the cores in the first place, the total
  star formation efficiency is only a few percent.\\
\end{enumerate}

The survey presented in this paper places important constraints on the
formation mechanisms, which should be taken into account in
theoretical studies of low-mass star formation. Naturally many
interesting problems remain which can be pursued using the results of
this paper as a starting point: for example a more detailed
characterization of the individual star forming cores (e.g., studying
variations in dynamics from molecular line widths and/or chemical
signatures) and of the embedded YSOs (e.g., their physical
properties/evolutionary stage by completing their SEDs from
near-infrared through (sub)mm wavelengths). It will also be
interesting to test whether the criteria from this paper can be used
to build comparable samples of starless and star-forming cores in
other nearby clouds. This will test whether the differing physical
conditions in different star-forming regions, such as the external
pressure, affects whether a given core forms stars or not. This might
for example reflect in a different value for the concentration above
which all submillimeter cores in a given region is found to have
embedded protostars. This paper will in that sense serve as a starting
point for such studies and for similar censuses of other clouds
surveyed at mid-infrared and (sub)millimeter wavelengths.

\acknowledgments We are grateful to Neal Evans, Melissa Enoch and
Luisa Rebull for useful discussions and comments about the paper. We
thank the referee for detailed comments which improved the
presentation of the results. The research of J.K.J. is supported by
NASA Origins Grant NAG5-13050. Support for this work, part of the
Spitzer Legacy Science Program, was also provided by NASA through
contract 1224608 issued by the Jet Propulsion Laboratory, California
Institute of Technology, under NASA contract 1407. D.J. is supported
by a Natural Sciences and Engineering Research Council of Canada
grant. H.K. is supported by a University of Victoria fellowship and a
National Research Council of Canada GSSSP award. This paper has made
use of the SIMBAD database, operated at CDS, Strasbourg, France.


\begin{thebibliography}{}

\bibitem[\protect\citeauthoryear{{Allen} et~al.}{{Allen}
  et~al.}{2004}]{allen04}
{Allen}, L.~E., et~al. 2004, \apjs, 154, 363

\bibitem[\protect\citeauthoryear{{Alves} \& {Lombardi}}{{Alves} \&
  {Lombardi}}{2006}]{alves06}
{Alves}, J.,  \& {Lombardi}, M. 2006, \apj, {in prep.}

\bibitem[\protect\citeauthoryear{{Andr\'e}, {Ward-Thompson}, \&
  {Barsony}}{{Andr\'e} et~al.}{1993}]{andre93}
{Andr\'e}, P., {Ward-Thompson}, D.,  \& {Barsony}, M. 1993, \apj, 406, 122

\bibitem[\protect\citeauthoryear{{Andr\'e}, {Ward-Thompson}, \&
  {Barsony}}{{Andr\'e} et~al.}{2000}]{andreppiv}
{Andr\'e}, P., {Ward-Thompson}, D.,  \& {Barsony}, M. 2000, in Protostars and
  Planets IV, ed. V. Mannings, A.\ P. Boss, \& S.\ S. Russell (University of
  Arizona Press, Tucson), 59

\bibitem[\protect\citeauthoryear{{Ayliffe} et~al.}{{Ayliffe}
  et~al.}{2006}]{ayliffe06}
{Ayliffe}, B.~A., {Langdon}, J.~C., {Cohl}, H.~S., \& {Bate}, M.~R. 2006, \mnras, {submitted}

\bibitem[\protect\citeauthoryear{{Ballesteros-Paredes}
  et~al.}{{Ballesteros-Paredes} et~al.}{2006}]{ballesterosppv}
{Ballesteros-Paredes}, J., {Klessen}, R., {Mac Low}, M.-M.,  \&
  {V\'{a}zquez-Semadeni}, E. 2006, in Protostars and Planets V, ed. B.
  Reipurth, D. Jewitt, \& K. Keil (University of Arizona Press, Tucson)

\bibitem[\protect\citeauthoryear{{Bate} et~al.}{{Bate}
  et~al.}{2003}]{bate03}
{Bate}, M.~R., {Bonnell}, I.~A., \& {Bromm}, V., 2003, \mnras, 339, 577

\bibitem[\protect\citeauthoryear{{Bondi} \& {Hoyle}}{{Bondi} \&
  {Hoyle}}{1944}]{bondi44}
{Bondi}, H.,  \& {Hoyle}, F. 1944, \mnras, 104, 273

\bibitem[\protect\citeauthoryear{{Bonnell} et~al.}{{Bonnell}
  et~al.}{1997}]{bonnell97}
{Bonnell}, I.~A., {Bate}, M.~R., {Clarke}, C.~J.,  \& {Pringle}, J.~E. 1997,
  \mnras, 285, 201

\bibitem[\protect\citeauthoryear{{Bonnell} et~al.}{{Bonnell}
  et~al.}{2001}]{bonnell01a}
{Bonnell}, I.~A., {Clarke}, C.~J., {Bate}, M.~R.,  \& {Pringle}, J.~E. 2001,
  \mnras, 324, 573

\bibitem[\protect\citeauthoryear{{Ciolek} \& {Basu}}{{Ciolek} \&
  {Basu}}{2001}]{ciolek01}
{Ciolek}, G.~E.,  \& {Basu}, S. 2001, \apj, 547, 272

\bibitem[\protect\citeauthoryear{{Crapsi} et~al.}{{Crapsi}
  et~al.}{2005}]{crapsi05}
{Crapsi}, A., {Caselli}, P., {Walmsley}, C.~M., {Myers}, P.~C., {Tafalla}, M.,
  {Lee}, C.~W.,  \& {Bourke}, T.~L. 2005, \apj, 619, 379

\bibitem[\protect\citeauthoryear{{Di Francesco} et~al.}{{Di Francesco}
  et~al.}{2006}]{difrancescoppv}
{Di Francesco}, J., {Evans}, N.~J.~I., {Caselli}, P., {Myers}, P.~C.,
  {Shirley}, Y., {Aikawa}, Y.,  \& {Tafalla}, M. 2006, in Protostars and
  Planets V, ed. B. Reipurth, D. Jewitt, \& K. Keil (University of Arizona
  Press, Tucson)

\bibitem[\protect\citeauthoryear{{Enoch} et~al.}{{Enoch}
  et~al.}{2006}]{enoch06}
{Enoch}, M.~L., et~al. 2006, \apj, 638, 293

\bibitem[\protect\citeauthoryear{{Evans} et~al.}{{Evans}
  et~al.}{2003}]{evans03}
{Evans}, N.~J., et~al. 2003, \pasp, 115, 965

\bibitem[\protect\citeauthoryear{{Evans} et~al.}{{Evans}
  et~al.}{2005}]{delivery3}
{Evans}, N.~J., et~al. {2005}, "Delivery of Data from the c2d Legacy Project:
  IRAC and MIPS (Pasadena, SSC)"

\bibitem[\protect\citeauthoryear{{Evans} et~al.}{{Evans}
  et~al.}{2001}]{evans01}
{Evans}, N.~J., {Rawlings}, J.~M.~C., {Shirley}, Y.~L.,  \& {Mundy}, L.~G.
  2001, \apj, 557, 193

\bibitem[\protect\citeauthoryear{{Goodman} et~al.}{{Goodman}
  et~al.}{2004}]{goodman04}
{Goodman}, A.~A., et~al. 2004, {ASP Conference Series}

\bibitem[\protect\citeauthoryear{{Greene} et~al.}{{Greene}
  et~al.}{1994}]{greene94}
{Greene}, T.~P., {Wilking}, B.~A., {Andre}, P., {Young}, E.~T.,  \& {Lada},
  C.~J. 1994, \apj, 434, 614

\bibitem[\protect\citeauthoryear{{Hartmann}}{{Hartmann}}{2002}]{hartmann02}
{Hartmann}, L. 2002, \apj, 578, 914

\bibitem[\protect\citeauthoryear{{Hatchell} et~al.}{{Hatchell}
  et~al.}{2005}]{hatchell05}
{Hatchell}, J., {Richer}, J.~S., {Fuller}, G.~A., {Qualtrough}, C.~J., {Ladd},
  E.~F.,  \& {Chandler}, C.~J. 2005, \aap, 440, 151

\bibitem[\protect\citeauthoryear{{Hogerheijde} \& {Sandell}}{{Hogerheijde} \&
  {Sandell}}{2000}]{hogerheijde00sandell}
{Hogerheijde}, M.~R.,  \& {Sandell}, G.~. 2000, \apj, 534, 880

\bibitem[\protect\citeauthoryear{{Huard} et~al.}{{Huard}
  et~al.}{2006}]{huard06}
{Huard}, T., et~al. {2006}, \apj, {in prep.}

\bibitem[\protect\citeauthoryear{{Jijina}, {Myers}, \& {Adams}}{{Jijina}
  et~al.}{1999}]{jijina99}
{Jijina}, J., {Myers}, P.~C.,  \& {Adams}, F.~C. 1999, \apjs, 125, 161

\bibitem[\protect\citeauthoryear{{Johnstone} \& {Bally}}{{Johnstone} \&
  {Bally}}{2006}]{johnstone06}
{Johnstone}, D.,  \& {Bally}, J. 2006, \apj, {in press. (astro-ph/0609171)}

\bibitem[\protect\citeauthoryear{{Johnstone}, {Di Francesco}, \&
  {Kirk}}{{Johnstone} et~al.}{2004}]{johnstone04}
{Johnstone}, D., {Di Francesco}, J.,  \& {Kirk}, H. 2004, \apjl, 611, L45

\bibitem[\protect\citeauthoryear{{Johnstone} et~al.}{{Johnstone}
  et~al.}{2000}]{johnstone00}
{Johnstone}, D., {Wilson}, C.~D., {Moriarty-Schieven}, G., {Joncas}, G.,
  {Smith}, G., {Gregersen}, E.,  \& {Fich}, M. 2000, \apj, 545, 327

\bibitem[\protect\citeauthoryear{{J{\o}rgensen} et~al.}{{J{\o}rgensen}
  et~al.}{2006}]{perspitz}
{J{\o}rgensen}, J.~K., et~al. 2006, \apj, 645, 1246

\bibitem[\protect\citeauthoryear{{J{\o}rgensen} et~al.}{{J{\o}rgensen}
  et~al.}{2005}]{iras16293letter}
{J{\o}rgensen}, J.~K., et~al. 2005, \apjl, 631, L77

\bibitem[\protect\citeauthoryear{{J{\o}rgensen}, {Sch{\" o}ier}, \& {van
  Dishoeck}}{{J{\o}rgensen} et~al.}{2002}]{jorgensen02}
{J{\o}rgensen}, J.~K., {Sch{\" o}ier}, F.~L.,  \& {van Dishoeck}, E.~F. 2002,
  \aap, 389, 908

\bibitem[\protect\citeauthoryear{{J{\o}rgensen}, {Sch\"{o}ier}, \& {van
  Dishoeck}}{{J{\o}rgensen} et~al.}{2005}]{coevollet}
{J{\o}rgensen}, J.~K., {Sch\"{o}ier}, F.~L.,  \& {van Dishoeck}, E.~F. 2005,
  \aap, 435, 177

\bibitem[\protect\citeauthoryear{{Kirk}, {Johnstone}, \& {Di Francesco}}{{Kirk}
  et~al.}{2006}]{kirk06}
{Kirk}, H., {Johnstone}, D.,  \& {Di Francesco}, J. 2006, \apj, 646, 1009

\bibitem[\protect\citeauthoryear{{Kirk}, {Ward-Thompson}, \&
  {Andr{\'e}}}{{Kirk} et~al.}{2005}]{kirk05}
{Kirk}, J.~M., {Ward-Thompson}, D.,  \& {Andr{\'e}}, P. 2005, \mnras, 360, 1506

\bibitem[\protect\citeauthoryear{{Lada}}{{Lada}}{1987}]{lada87}
{Lada}, C.~J. 1987, in IAU Symp. 115: Star Forming Regions (D. Reidel
  Publishing Co., Dordrecht), Vol. 115, 1

\bibitem[\protect\citeauthoryear{{Lada} \& {Lada}}{{Lada} \&
  {Lada}}{2003}]{lada03}
{Lada}, C.~J.,  \& {Lada}, E.~A. 2003, \araa, 41, 57

\bibitem[\protect\citeauthoryear{{Ladd}, {Myers}, \& {Goodman}}{{Ladd}
  et~al.}{1994}]{ladd94}
{Ladd}, E.~F., {Myers}, P.~C.,  \& {Goodman}, A.~A. 1994, \apj, 433, 117

\bibitem[\protect\citeauthoryear{{Lee} \& {Myers}}{{Lee} \&
  {Myers}}{1999}]{lee99}
{Lee}, C.~W.,  \& {Myers}, P.~C. 1999, \apjs, 123, 233

\bibitem[\protect\citeauthoryear{{Motte}, {Andr\'e}, \& {Neri}}{{Motte}
  et~al.}{1998}]{motte98}
{Motte}, F., {Andr\'e}, P.,  \& {Neri}, R. 1998, \aap, 336, 150

\bibitem[\protect\citeauthoryear{{Muzerolle} et~al.}{{Muzerolle}
  et~al.}{2004}]{muzerolle04}
{Muzerolle}, J., et~al. 2004, \apjs, 154, 379

\bibitem[\protect\citeauthoryear{{Ossenkopf} \& {Henning}}{{Ossenkopf} \&
  {Henning}}{1994}]{ossenkopf94}
{Ossenkopf}, V.,  \& {Henning}, T. 1994, \aap, 291, 943

\bibitem[\protect\citeauthoryear{{Rebull} et~al.}{{Rebull}
  et~al.}{2006}]{rebull06}
{Rebull}, L., et~al. {2006}, \apj, {submitted}

\bibitem[\protect\citeauthoryear{{Ridge} et~al.}{{Ridge}
  et~al.}{2006}]{ridge06}
{Ridge}, N.~A., et~al. 2006, \aj, 131, 2921

\bibitem[\protect\citeauthoryear{{Shirley}, {Evans}, \& {Rawlings}}{{Shirley}
  et~al.}{2002}]{shirley02}
{Shirley}, Y.~L., {Evans}, N.~J.,  \& {Rawlings}, J.~M.~C. 2002, \apj, 575, 337

\bibitem[\protect\citeauthoryear{{Tapia} et~al.}{{Tapia}
  et~al.}{1997}]{tapia97}
{Tapia}, M., {Persi}, P., {Bohigas}, J.,  \& {Ferrari-Toniolo}, M. 1997, \aj,
  113, 1769

\bibitem[\protect\citeauthoryear{{Walawender} et~al.}{{Walawender}
  et~al.}{2005}]{walawender05b1}
{Walawender}, J., {Bally}, J., {Kirk}, H.,  \& {Johnstone}, D. 2005, \aj, 130,
  1795

\bibitem[\protect\citeauthoryear{{Walawender} et~al.}{{Walawender}
  et~al.}{2006}]{walawender06}
{Walawender}, J., {Bally}, J., {Kirk}, H., {Johnstone}, D., {Reipurth}, B., \& Aspin, C. 2006, \aj, 132, 467 

\bibitem[\protect\citeauthoryear{{Walsh}, {Myers}, \& {Burton}}{{Walsh}
  et~al.}{2004}]{walsh04}
{Walsh}, A.~J., {Myers}, P.~C.,  \& {Burton}, M.~G. 2004, \apj, 614, 194

\bibitem[\protect\citeauthoryear{{Ward-Thompson} et~al.}{{Ward-Thompson}
  et~al.}{2006}]{wardthompsonppv}
{Ward-Thompson}, D., {Andr\'{e}}, P., {Crutcher}, R., {Johnstone}, D.,
  {Onishi}, T.,  \& {Wilson}, C. 2006, in Protostars and Planets V, ed. B.
  Reipurth, D. Jewitt, \& K. Keil (University of Arizona Press, Tucson)

\bibitem[\protect\citeauthoryear{{Whitney} et~al.}{{Whitney}
  et~al.}{2003}]{whitney03b}
{Whitney}, B.~A., {Wood}, K., {Bjorkman}, J.~E.,  \& {Cohen}, M. 2003, \apj,
  598, 1079

\bibitem[\protect\citeauthoryear{{Williams}, {de Geus}, \& {Blitz}}{{Williams}
  et~al.}{1994}]{williams94}
{Williams}, J.~P., {de Geus}, E.~J.,  \& {Blitz}, L. 1994, \apj, 428, 693

\bibitem[\protect\citeauthoryear{{Young} et~al.}{{Young}
  et~al.}{2006}]{young06}
{Young}, K.~E., et~al. 2006, \apj, 644, 326

\end{thebibliography}
\end{document}